\documentclass[letterpaper, 10 pt, conference]{ieeeconf}  

\usepackage{latexsym,amssymb,amsmath,ifthen,afterpage,cite,amsfonts}
\usepackage{color}
\usepackage{algorithm}
\usepackage{algorithmic}
\usepackage{enumerate}
\usepackage{enumerate}
\usepackage{subfig}

\ifx\pdfoutput\undefined
\usepackage{graphicx}
\graphicspath{{eps_figs/}}
\else
\usepackage[pdftex]{graphicx}
\graphicspath{{pdf_figs/}}
\fi

\usepackage{tikz} 
\usepackage{pgfplots}
\usepackage{pgfplotstable}
\usepackage{filecontents}

\newcommand{\overbar}[1]{\mkern 1.5mu\overline{\mkern-1.5mu#1\mkern-1.5mu}\mkern 1.5mu}

\newcommand{\Fig}[1]{Fig.~\ref{#1}}

\newcommand{\eqdef}{\stackrel{\scriptscriptstyle\bigtriangleup}{=} }

\newcommand{\R}{\mathbb{R}}

\newcommand{\calX}{\mathcal{X}}

\newcommand{\calF}{\mathcal{F}}

\newcommand{\x}{{\bf x}}
\newcommand{\y}{{\bf y}}
\newcommand{\Y}{{\bf Y}}

\newcommand{\E}{\mathop{\mathbb{E}}}
\newcommand{\V}{\mathop{\mathbb{V}}}
\newcommand{\Var}{\mathop{\mathbb{V}}}

\newcounter{examplecntr}
{\begin{trivlist}\small\item[]\refstepcounter{examplecntr}%
 {\bfseries Example~\theexamplecntr%
  \ifthenelse{\equal{#1}{}}{}{ (#1)}.
}}%
{\end{trivlist}}

\newcounter{theoremcntr}
{\begin{trivlist}\item[]\refstepcounter{theoremcntr}%
{\bfseries Theorem~\thetheoremcntr%
  \ifthenelse{\equal{#1}{}}{}{ (#1)}.
}}%
{\hfill$\Box$\end{trivlist}}

\newcommand{\pushright}[1]{\ifmeasuring@#1\else\omit\hfill$\displaystyle#1$\fi\ignorespaces}

\newcommand{\pos}[2]{\makebox(0,0)[#1]{#2}}

  \renewenvironment{thebibliography}[1]{%
    \begin{oldthebibliography}{#1}%
      \setlength{\itemsep}{0.252ex}%
  }%
  {%
    \end{oldthebibliography}%
  }
  
\begin{document}
\DeclareGraphicsExtensions{.pdf}

\eject \pdfpagewidth = 8.5in
\eject \pdfpageheight = 11in

\title{The Primal versus the Dual Ising Model} 

\author{Mehdi Molkaraie \\
ETH Zurich \\
\tt{mehdi.molkaraie@alumni.ethz.ch}
}

\maketitle 
 
\begin{abstract}
We represent the Ising model of statistical physics 
by normal factor graphs in the primal and in the 
dual domains. By analogy with Kirchhoff's voltage and current laws, we show that in the primal 
normal factor graphs, the dependency among the variables is along the cycles, whereas in the dual normal 
factor graphs, the 
dependency is on the cutsets. In the primal (resp. dual) domain, dependent
variables can be computed via their fundamental cycles (resp. fundamental cutsets). 
Using Onsager's closed form solution, we illustrate the opposite 
behavior of the uniform sampling estimator for estimating the partition function in the primal and in the dual of the homogeneous Ising model on a two-dimensional torus. 
\end{abstract} 
 
\section{Introduction} 

We relate some properties of the normal/Forney factor graph (NFG)~\cite{Forney:01}  
representation of the Ising model~\cite{Cipra:87,Baxter07} in 
the primal and in the dual domains to pertinent results in algebraic graph theory. 
The focus is on Ising models with arbitrary topology, with pairwise (nearest-neighbor) interactions, 
and without an external magnetic field. 

Here, the NFG is a simple, finite, and connected graph \mbox{$G=(V,E)$}, where $V$ is the set of 
vertices and $E$ the set of edges. 
In our analysis, we use partitionings of $G$ into 
\mbox{$G = T \cup \overbar T$}, where $T$ is a spanning 
tree and $\overbar T$ is the corresponding cospanning tree. The edges of $T$ are called 
the \emph{branches} 
and the edges of $\overbar T$ are called the \emph{chords} of $G$ with respect to $T$, or simply
the chords of $T$. 
Since $G$ is connected, $|E| \ge |V| - 1$ and 
\begin{IEEEeqnarray}{rCl}
|T| & = & |V|-1 \label{eqn:SizeT}\\
|\overbar T| & = & |E| - |V| + 1, \label{eqn:SizeS}
\end{IEEEeqnarray}
where $|\cdot|$ denotes the cardinality of a set. 

We prove that the sum (modulo 2)
of variables along any cycle in the primal NFG is zero. 
In the dual NFG, we prove that the sum (modulo 2)
of variables on any cutset is zero. We then
propose a uniform sampling algorithm for estimating the partition function in the primal and 
in the dual domains.
For the homogeneous Ising model on a two-dimensional (2D) torus, we employ Onsager's 
analytical solution~\cite{Onsager:44, Baxter07} to illustrate the opposite behavior of
the uniform sampling estimator in the primal and in the dual domains.

For more details on the cycle 
space, the cutset space, and their duality in the context of algebraic 
graph theory, see~\cite[Chapter 2]{Bolob},~\cite[Chapter 14]{Godsil}.
The topic of the Kramers--Wannier duality (which relates the partition function of the 2D Ising model at high and at low temperatures) 
is discussed in the context of NFG duality 
in~\cite{AlVo:ISIT2015, AlVo:2016}.
Finally, we would like to point out that David Forney has recently developed and generalized some of the results of 
this paper  
in the context of algebraic topology. For more details, see~\cite{Forney:16}.

The paper is organized as follows. In Section~\ref{sec:Ising},
we review the Ising model and its graphical model representation 
in terms of NFGs. 
In Sections~\ref{sec:IS} and~\ref{sec:VarIsing}, we describe the uniform sampling 
algorithm in the primal domain and derive its variance for 
the Ising model on a 2D torus. The dual NFG of the model is discussed
in Section~\ref{sec:Dual}, and an analogous estimator based on uniform sampling in the dual NFG is 
presented in Section~\ref{sec:ISD}. The variance of the estimator in the dual NFG is discussed 
in Section~\ref{sec:VarIsingD}. The scale factor between the partition functions of 
the primal and the dual NFGs is derived in the Appendix. 

\section{The Ising Model in the Primal Domain}
\label{sec:Ising} 

Let ${\bf X} = (X_1, X_2, \ldots, X_N)$ be a collection of $N$ interacting discrete random 
variables taking values in a finite alphabet $\calX$, which in this context 
is equal to the binary field $\mathbb{F}_2$ and
let $x_i$ represents a possible realization of $X_i$. The vectors $\x \in \calX^N$ will be called
configurations.

The variables $X_1, X_2, \ldots, X_N$ are associated with the vertices of a graph
$G = (V, E)$ with $N$ vertices and $|E|$ edges. In the Ising model, each variable 
is assigned a spin, which represents the two possible states of a particle. 
Two variables interact if their 
corresponding vertices are connected by an edge in $G$.
Each edge has an associated coupling parameter $J_{k, \ell}$, which measures the strength of
the interaction between neighboring pair $(X_k, X_\ell)$.

Let $f \colon \calX^N \! \rightarrow \R_{\ge 0}$ be a non-negative real function which factors
into a product of local factors $\upsilon_{k, \ell}(\cdot)$ as
\begin{IEEEeqnarray}{c}
\label{eqn:factorF}
f(\x) = \prod_{\text{$(k,\ell) \in E$}}\upsilon_{k, \ell}(x_k, x_{\ell}),
\end{IEEEeqnarray}
where $E$ contains all the unordered distinct interacting pairs $(k, \ell)$, and
$\upsilon_{k, \ell} \colon \calX^2 \! \rightarrow \R_{\ge 0}$ is given by 
\begin{equation} 
\label{eqn:IsingA}
\upsilon_{k, \ell}(x_k, x_{\ell}) = \left\{ \begin{array}{ll}
     e^{\beta J_{k,\ell}}, & \text{if $x_k = x_{\ell}$} \\
     e^{-\beta J_{k,\ell}}, & \text{if $x_k \ne x_{\ell}$,}
  \end{array} \right.
\end{equation}
where $\beta$ is the inverse temperature. The Ising model is called ferromagnetic 
(resp.~antiferromagnetic) if $J_{k,\ell} > 0$ (resp. $J_{k,\ell} < 0$) for
all $(k, \ell) \in E$. 

We will find it convenient to set $\beta = 1$ and to work with varying values of the coupling 
parameters. 
In this setup, large values of $|J_{k,\ell}|$ correspond to the
low-temperature regime, and small values of $|J_{k,\ell}|$ correspond to 
the high-temperature regime. In particular, $J_{k,\ell}=0$ for all $(k,\ell)\in E$ 
corresponds to infinite temperature.

From~(\ref{eqn:factorF}), the Boltzmann distribution is defined as~\cite{Baxter07}
\begin{IEEEeqnarray}{c}
\label{eqn:Prob}
p\hspace{0.15mm}_{\text{B}}(\x) \eqdef \frac{f(\x)}{Z}\cdot    
\end{IEEEeqnarray}

Here, $Z$ is the \emph{partition function}, which makes $p\hspace{0.15mm}_{\text{B}}(\cdot)$ a 
probability mass function 
over $\calX^N$, and is given by
\begin{IEEEeqnarray}{c}
\label{eqn:PartFunction}
Z = \sum_{\x} f(\x), 
\end{IEEEeqnarray}
where the summation runs over all configurations.

The factorization in~(\ref{eqn:factorF}) can be 
represented
by an NFG $G=(V,E)$, in which vertices represent the factors 
and edges 
represent the variables.
The edge 
that represents some variable $x$ is connected to the vertex 
representing the factor $\upsilon(\cdot)$
if and only if $x$ is an argument of $\upsilon(\cdot)$. If a 
variable (an edge) appears in more than  two factors, such a variable is 
replicated using an equality indicator factor~\cite{Forney:01}. 

The primal NFGs of the Ising model on a chain (1D graph with periodic boundaries), a 2D lattice, and 
a fully-connected graph are shown in~\Fig{fig:2DGrid},
where the unlabeled 
boxes represent~(\ref{eqn:IsingA}) and boxes labeled ``$=$'' are equality 
indicator factors. 
For example, in~\Fig{fig:2DGrid}--left the equality indicator factor $\Phi_{=}(\cdot)$ involving 
variables $x_2, x'_2,$ and $x''_2$ is given by 
\begin{IEEEeqnarray}{c}
\label{eqn:equality}
\Phi_{=}(x_2, x'_2, x''_2) = \delta( x_2 - x'_2)\cdot\delta(x_2- x''_2),
\end{IEEEeqnarray}
where $\delta(\cdot)$ is the Kronecker delta function.

%


We note that each factor~(\ref{eqn:IsingA})
is only a function of \mbox{$x_k + x_{\ell}$}. (Recall that arithmetic manipulations 
are done modulo $2$.) We can thus 
represent $\upsilon_{k, \ell}(\cdot)$ using 
only one variable $y_e$, as
\begin{equation} 
\label{eqn:IsingKernelOrig}
\upsilon_{e}(y_e) = \left\{ \begin{array}{ll}
     e^{J_e}, & \text{if $y_e = 0$} \\
     e^{-J_e}, & \text{if $y_e = 1$.}
  \end{array} \right.
\end{equation}

Let $\Y$ denote $(Y_1, Y_2, \ldots, Y_{|E|})$, where $|\Y| = |E|$.








\begin{figure}[t!!]
\setlength{\unitlength}{0.78mm}
\centering
\begin{picture}(103,66)(0,0)
 \linethickness{0.2mm}
\small
%
\put(32,58){\framebox(4,4){$=$}}
\put(36,60){\line(1,0){8}}        
\put(44,58){\framebox(4,4){}}
\put(48,60){\line(1,0){8}}
\put(56,58){\framebox(4,4){$=$}}  
\put(60,60){\line(1,0){8}}      
\put(68,58){\framebox(4,4){}}
\put(72,60){\line(1,0){8}}
\put(80,58){\framebox(4,4){$=$}}
\put(34,53){\line(0,1){5}}
\put(34,53){\line(1,0){22}}
\put(56,51){\framebox(4,4){}}
\put(82,53){\line(0,1){5}}
\put(60,53){\line(1,0){22}}
%
%
%
 \put(9,42.8){\pos{bc}{$X_1$}}
 \put(21,42.8){\pos{bc}{$X_2$}} 
 \put(30.26,35.1){\pos{bc}{$X''_2$}}
 \put(33.2,42.8){\pos{bc}{$X'_2$}}
\put(1,40){\framebox(4,4){$=$}}
\put(5,42){\line(1,0){8}}
\put(13,40){\framebox(4,4){}}
\put(17,42){\line(1,0){8}}
\put(25,40){\framebox(4,4){$=$}}
\put(29,42){\line(1,0){8}}
\put(37,40){\framebox(4,4){}}
\put(41,42){\line(1,0){8}}
\put(49,40){\framebox(4,4){$=$}}
\put(3,34){\line(0,1){6}}
\put(1,30){\framebox(4,4){}}
\put(3,30){\line(0,-1){6}}
\put(27,34){\line(0,1){6}}
\put(25,30){\framebox(4,4){}}
\put(27,30){\line(0,-1){6}}
\put(51,34){\line(0,1){6}}
\put(49,30){\framebox(4,4){}}
\put(51,30){\line(0,-1){6}}
\put(1,20){\framebox(4,4){$=$}}
\put(5,22){\line(1,0){8}}
\put(13,20){\framebox(4,4){}}
\put(17,22){\line(1,0){8}}
\put(25,20){\framebox(4,4){$=$}}
\put(29,22){\line(1,0){8}}
\put(37,20){\framebox(4,4){}}
\put(41,22){\line(1,0){8}}
\put(49,20){\framebox(4,4){$=$}}
\put(3,14){\line(0,1){6}}
\put(1,10){\framebox(4,4){}}
\put(3,10){\line(0,-1){6}}
\put(27,14){\line(0,1){6}}
\put(25,10){\framebox(4,4){}}
\put(27,10){\line(0,-1){6}}
\put(51,14){\line(0,1){6}}
\put(49,10){\framebox(4,4){}}
\put(51,10){\line(0,-1){6}}
\put(1,0){\framebox(4,4){$=$}}
\put(5,2){\line(1,0){8}}
\put(13,0){\framebox(4,4){}}
\put(17,2){\line(1,0){8}}
\put(25,0){\framebox(4,4){$=$}}
\put(29,2){\line(1,0){8}}
\put(37,0){\framebox(4,4){}}
\put(41,2){\line(1,0){8}}
\put(49,0){\framebox(4,4){$=$}}
%
%
\put(69,0){\framebox(4,4){$=$}}
\put(73,2){\line(1,0){8}}
\put(81,0){\framebox(4,4){}}
\put(85,2){\line(1,0){8}}
\put(93,0){\framebox(4,4){$=$}}
\put(95.325,4.05){\rotatebox{72}{\line(1,0){7.6}}}
\put(96.6,11.4){\framebox(4,4){}}
\put(99.2,15.5){\rotatebox{72}{\line(1,0){7.5}}}
\put(100.416,22.825){\framebox(4,4){$=$}}
\put(101.675,26.925){\rotatebox{-36}{\line(-1,0){7.6}}}
\put(91.5,30.9){\framebox(4,4){}}
\put(91.3,34.5){\rotatebox{-36}{\line(-1,0){7.65}}}
\put(64.6,26.925){\rotatebox{37}{\line(1,0){7.25}}}
\put(70.5,30.9){\framebox(4,4){}}
\put(74.5,34.4){\rotatebox{37}{\line(1,0){7.8}}}
\put(70.675,4.05){\rotatebox{-72}{\line(-1,0){7.6}}}
\put(65.6,11.4){\framebox(4,4){}}
\put(67,15.5){\rotatebox{-72}{\line(-1,0){7.5}}}
\put(61.684,22.825){\framebox(4,4){$=$}}
\put(81,37.932){\framebox(4,4){$=$}}
\put(65.7,24.825){\line(1,0){15.1}}
\put(100.4,24.825){\line(-1,0){15.2}}
\put(81,22.825){\framebox(4,4){}}
\put(93.675,4.05){\rotatebox{-35}{\line(-1,0){14.3}}}
\put(65.8,23.55){\rotatebox{-36}{\line(1,0){14.7}}}
\put(77.9,11.4){\framebox(4,4){}}
\put(72.3,4.05){\rotatebox{35}{\line(1,0){14.25}}}
\put(84.2,11.4){\framebox(4,4){}}
\put(100.2,23.55){\rotatebox{36}{\line(-1,0){14.75}}}
\put(71.1,4.05){\rotatebox{72}{\line(1,0){15.4}}}
\put(82.1,37.9){\rotatebox{72}{\line(-1,0){15.7}}}
\put(74,18.8){\framebox(4,4){}}
\put(95,4.05){\rotatebox{-72}{\line(-1,0){15.4}}}
\put(83.95,37.7){\rotatebox{-72}{\line(1,0){15.7}}}
\put(88,18.8){\framebox(4,4){}}
\end{picture}
\vspace{1.0ex}
\caption{\label{fig:2DGrid}%
The Primal NFG of the Ising model on a (top) chain (left) 2D lattice 
(right) fully-connected graph.
The unlabeled 
boxes represent~(\ref{eqn:IsingA}) 
and boxes containing $``="$ symbols are given by~(\ref{eqn:equality}).}
\vspace{4.0ex}
\centering
\begin{picture}(103,64)(0,0)
 \linethickness{0.2mm}
 \setlength{\unitlength}{0.76mm}
\small
%
\put(32,62){\framebox(4,4){$=$}}
\put(44,62){\framebox(4,4){$+$}}

\put(56,62){\framebox(4,4){$=$}}  
\put(68,62){\framebox(4,4){$+$}}
\put(80,62){\framebox(4,4){$=$}}
\put(34,57){\line(0,1){5}}
\put(34,57){\line(1,0){22}}
\put(56,55){\framebox(4,4){$+$}}
\put(82,57){\line(0,1){5}}
\put(60,57){\line(1,0){22}}
%
%
\put(6,40){\framebox(4,4){$=$}}
\put(18,40){\framebox(4,4){$+$}}
\put(30,40){\framebox(4,4){$=$}}
\put(34,42){\line(1,0){8}}
\put(42,40){\framebox(4,4){$+$}}
\put(46,42){\line(1,0){8}}
\put(54,40){\framebox(4,4){$=$}}
\put(8,34){\line(0,1){6}}
\put(6,30){\framebox(4,4){$+$}}
\put(8,30){\line(0,-1){6}}
\put(30,30){\framebox(4,4){$+$}}
\put(54,30){\framebox(4,4){$+$}}
%
\put(6,20){\framebox(4,4){$=$}}
\put(18,20){\framebox(4,4){$+$}}
\put(30,20){\framebox(4,4){$=$}}
\put(42,20){\framebox(4,4){$+$}}
\put(54,20){\framebox(4,4){$=$}}
%
\put(6,10){\framebox(4,4){$+$}}
\put(32,14){\line(0,1){6}}
\put(30,10){\framebox(4,4){$+$}}
\put(32,10){\line(0,-1){6}}
\put(54,10){\framebox(4,4){$+$}}
%
\put(6,0){\framebox(4,4){$=$}}
\put(18,0){\framebox(4,4){$+$}}
\put(30,0){\framebox(4,4){$=$}}
\put(34,2){\line(1,0){8}}
\put(42,0){\framebox(4,4){$+$}}
\put(46,2){\line(1,0){8}}
\put(54,0){\framebox(4,4){$=$}}
%
%
\put(73,6){\framebox(4,4){$=$}}
\put(85,6){\framebox(4,4){$+$}}
\put(97,6){\framebox(4,4){$=$}}
%
\put(100.6,17.4){\framebox(4,4){$+$}}
\put(104.416,28.825){\framebox(4,4){$=$}}
%
\put(95.5,36.9){\framebox(4,4){$+$}}
%
\put(68.6,32.925){\rotatebox{37}{\line(1,0){7.25}}}
\put(74.5,36.9){\framebox(4,4){$+$}}
\put(78.5,40.4){\rotatebox{37}{\line(1,0){7.8}}}
\put(74.675,10.05){\rotatebox{-72}{\line(-1,0){7.6}}}
\put(69.6,17.4){\framebox(4,4){$+$}}
\put(70,21.5){\rotatebox{-72}{\line(-1,0){7.5}}}
\put(65.684,28.825){\framebox(4,4){$=$}}
\put(85,43.932){\framebox(4,4){$=$}}
%
\put(85,28.825){\framebox(4,4){$+$}}
\put(97.675,10.05){\rotatebox{-35}{\line(-1,0){14.3}}}
\put(69.8,29.55){\rotatebox{-36}{\line(1,0){14.7}}}
\put(81.9,17.4){\framebox(4,4){$+$}}
\put(76.3,10.05){\rotatebox{35}{\line(1,0){14.25}}}
\put(88.2,17.4){\framebox(4,4){$+$}}
\put(104.2,29.55){\rotatebox{36}{\line(-1,0){14.75}}}
\put(75.1,10.05){\rotatebox{72}{\line(1,0){15.4}}}
\put(86.1,43.9){\rotatebox{72}{\line(-1,0){15.7}}}
\put(78,24.8){\framebox(4,4){$+$}}
\put(99,10.05){\rotatebox{-72}{\line(-1,0){15.4}}}
\put(87.95,43.7){\rotatebox{-72}{\line(1,0){15.7}}}
\put(92,24.8){\framebox(4,4){$+$}}
%
\put(46,66){\line(0,1){2}}
\put(44,68){\framebox(4,4){}}
\put(70,66){\line(0,1){2}}
\put(68,68){\framebox(4,4){}}
\put(58,55){\line(0,-1){2}}
\put(56,49){\framebox(4,4){}}
%
\put(20,44){\line(0,1){2}}
\put(44,44){\line(0,1){2}}
\put(18,46){\framebox(4,4){}}
\put(42,46){\framebox(4,4){}}
\put(6,32){\line(-1,0){2}}
\put(30,32){\line(-1,0){2}}
\put(54,32){\line(-1,0){2}}
\put(0,30){\framebox(4,4){}}
\put(24,30){\framebox(4,4){}}
\put(48,30){\framebox(4,4){}}
\put(20,24){\line(0,1){2}}
\put(44,24){\line(0,1){2}}
\put(18,26){\framebox(4,4){}}
\put(42,26){\framebox(4,4){}}
\put(6,12){\line(-1,0){2}}
\put(30,12){\line(-1,0){2}}
\put(54,12){\line(-1,0){2}}
\put(0,10){\framebox(4,4){}}
\put(24,10){\framebox(4,4){}}
\put(48,10){\framebox(4,4){}}
\put(20,4){\line(0,1){2}}
\put(44,4){\line(0,1){2}}
\put(18,6){\framebox(4,4){}}
\put(42,6){\framebox(4,4){}}
%
%
\put(87,6){\line(0,-1){2}}
\put(85,0){\framebox(4,4){}}
\put(104.6,19.4){\line(1,0){2}}
\put(106.6,17.4){\framebox(4,4){}}
\put(99.5,38.9){\line(1,0){2.2}}
\put(101.7,36.9){\framebox(4,4){}}
\put(69.6,19.4){\line(-1,0){2}}
\put(63.6,17.4){\framebox(4,4){}}
\put(87.02,32.825){\line(0,1){2}}
\put(85.10,34.825){\framebox(3.8,3.8){}}
\put(79.9,19.4){\line(1,0){2}}
\put(75.9,17.4){\framebox(4,4){}}
\put(92.2,19.4){\line(1,0){2}}
\put(94.2,17.4){\framebox(4,4){}}
\put(78,26.8){\line(-1,0){2}}
\put(72,24.8){\framebox(4,4){}}
\put(96,26.8){\line(1,0){2}}
\put(98,24.8){\framebox(4,4){}}
\put(72.5,38.9){\line(1,0){2}}
\put(68.5,36.9){\framebox(4,4){}}

\linethickness{0.48mm}
%
\put(36,64){\line(1,0){8}}        
\put(48,64){\line(1,0){8}}
\put(60,64){\line(1,0){8}}      
\put(72,64){\line(1,0){8}}
 %
\put(10,42){\line(1,0){8}}
\put(22,42){\line(1,0){8}}
\put(32,34){\line(0,1){6}}
\put(32,30){\line(0,-1){6}}
\put(56,34){\line(0,1){6}}
\put(56,30){\line(0,-1){6}}
\put(10,22){\line(1,0){8}}
\put(22,22){\line(1,0){8}}
\put(34,22){\line(1,0){8}}
\put(46,22){\line(1,0){8}}
\put(8,14){\line(0,1){6}}
\put(8,10){\line(0,-1){6}}
\put(56,14){\line(0,1){6}}
\put(56,10){\line(0,-1){6}}
\put(10,2){\line(1,0){8}}
\put(22,2){\line(1,0){8}}
%
\put(77,8){\line(1,0){8}}
\put(89,8){\line(1,0){8}}
\put(99.325,10.05){\rotatebox{72}{\line(1,0){7.6}}}
\put(103.2,21.5){\rotatebox{72}{\line(1,0){7.5}}}
\put(105.675,32.99){\rotatebox{-36}{\line(-1,0){7.67}}}
\put(95.2,40.2){\rotatebox{-36}{\line(-1,0){7.71}}}
\put(69.7,30.825){\line(1,0){15.1}}
\put(104.4,30.825){\line(-1,0){15.2}}
%
\put(24.75,44.6){\pos{bc}{$Y_1$}}
\put(14,42.8){\pos{bc}{$X_1$}}
 \put(26,38.0){\pos{bc}{$X_2$}} 
\end{picture}
\vspace{1.3ex}
\caption{\label{fig:2DGridMod}%
Modified primal NFGs of the Ising models on a (top) chain (left) 2D lattice 
(right) fully-connected graph..
The unlabeled boxes represent~(\ref{eqn:IsingKernelOrig}), boxes containing 
$``+"$ symbols are as in~(\ref{eqn:XOR}), and boxes containing $``="$ symbols 
are given by~(\ref{eqn:equality}). In each NFG, the branches of a spanning tree are 
marked by thick black edges.}
\end{figure}


Following the above observation, we construct the ``modified" primal NFGs of the Ising 
models illustrated
in~\Fig{fig:2DGridMod}, where the 
unlabeled boxes represent~(\ref{eqn:IsingKernelOrig}) and 
boxes labeled ``$+$'' are zero-sum indicator factors, which impose the
constraint that all their incident variables sum 
to zero. 
For example, in~\Fig{fig:2DGridMod}--left the zero-sum indicator factor $\Phi_{+}(\cdot)$ involving 
$x_1, x_2,$ and $y_1$ is given by
\begin{IEEEeqnarray}{c}
\label{eqn:XOR}
\Phi_{+}(y_1, x_1,  x_2)  =   \delta (y_1 + x_1 + x_2).
\end{IEEEeqnarray}

In the sequel, we drop the adjective ``modified" and 
refer to the NFGs in~\Fig{fig:2DGridMod} as 
the primal NFGs of the Ising model, when it causes no confusion.





\begin{figure}[t]
\setlength{\unitlength}{0.72mm}
\centering
\begin{picture}(77,72.5)(0,0)
 \linethickness{0.2mm}
\small
\put(0,60){\framebox(4,4){$=$}}
\put(4,62){\line(1,0){8}}        
\put(12,60){\framebox(4,4){$+$}}  
\put(16,62){\line(1,0){8}} 
\put(24,60){\framebox(4,4){$=$}}
\put(28,62){\line(1,0){8}}       
\put(36,60){\framebox(4,4){$+$}}
\put(40,62){\line(1,0){8}}
\put(48,60){\framebox(4,4){$=$}}
\put(52,62){\line(1,0){8}}       
\put(60,60){\framebox(4,4){$+$}} 
\put(64,62){\line(1,0){8}}
\put(72,60){\framebox(4,4){$=$}}
%
\put(2,54){\line(0,1){6}}
\put(0,50){\framebox(4,4){$+$}}
\put(2,50){\line(0,-1){6}}
\put(26,54){\line(0,1){6}}
\put(24,50){\framebox(4,4){$+$}}
\put(26,50){\line(0,-1){6}}
\put(50,54){\line(0,1){6}}
\put(48,50){\framebox(4,4){$+$}}
\put(50,50){\line(0,-1){6}}
\put(74,54){\line(0,1){6}}
\put(72,50){\framebox(4,4){$+$}}
\put(74,50){\line(0,-1){6}}
\put(0,40){\framebox(4,4){$=$}}
\put(4,42){\line(1,0){8}}
\put(12,40){\framebox(4,4){$+$}}
\put(16,42){\line(1,0){8}}
\put(24,40){\framebox(4,4){$=$}}
\put(28,42){\line(1,0){8}}
\put(36,40){\framebox(4,4){$+$}}
\put(40,42){\line(1,0){8}}
\put(48,40){\framebox(4,4){$=$}}
\put(52,42){\line(1,0){8}}
\put(60,40){\framebox(4,4){$+$}}
\put(64,42){\line(1,0){8}}
\put(72,40){\framebox(4,4){$=$}}
%
\put(2,34){\line(0,1){6}}
\put(0,30){\framebox(4,4){$+$}}
\put(2,30){\line(0,-1){6}}
\put(26,34){\line(0,1){6}}
\put(24,30){\framebox(4,4){$+$}}
\put(26,30){\line(0,-1){6}}
\put(50,34){\line(0,1){6}}
\put(48,30){\framebox(4,4){$+$}}
\put(50,30){\line(0,-1){6}}
\put(74,34){\line(0,1){6}}
\put(72,30){\framebox(4,4){$+$}}
\put(74,30){\line(0,-1){6}}
\put(0,20){\framebox(4,4){$=$}}
\put(4,22){\line(1,0){8}}
\put(12,20){\framebox(4,4){$+$}}
\put(16,22){\line(1,0){8}}
\put(24,20){\framebox(4,4){$=$}}
\put(28,22){\line(1,0){8}}
\put(36,20){\framebox(4,4){$+$}}
\put(40,22){\line(1,0){8}}
\put(48,20){\framebox(4,4){$=$}}
\put(52,22){\line(1,0){8}}
\put(60,20){\framebox(4,4){$+$}}
\put(64,22){\line(1,0){8}}
\put(72,20){\framebox(4,4){$=$}}
%
\put(2,14){\line(0,1){6}}
\put(0,10){\framebox(4,4){$+$}}
\put(2,10){\line(0,-1){6}}
\put(26,14){\line(0,1){6}}
\put(24,10){\framebox(4,4){$+$}}
\put(26,10){\line(0,-1){6}}
\put(50,14){\line(0,1){6}}
\put(48,10){\framebox(4,4){$+$}}
\put(50,10){\line(0,-1){6}}
\put(74,14){\line(0,1){6}}
\put(72,10){\framebox(4,4){$+$}}
\put(74,10){\line(0,-1){6}}
\put(0,0){\framebox(4,4){$=$}}
\put(4,2){\line(1,0){8}}
\put(12,0){\framebox(4,4){$+$}}
\put(16,2){\line(1,0){8}}
\put(24,0){\framebox(4,4){$=$}}
\put(28,2){\line(1,0){8}}
\put(36,0){\framebox(4,4){$+$}}
\put(40,2){\line(1,0){8}}
\put(48,0){\framebox(4,4){$=$}}
\put(52,2){\line(1,0){8}}
\put(60,0){\framebox(4,4){$+$}}
\put(64,2){\line(1,0){8}}
\put(72,0){\framebox(4,4){$=$}}
 
\put(19,64.1){\pos{bc}{$Y_1$}}
\put(43,64.1){\pos{bc}{$Y_2$}}
\put(62,64){\line(0,1){2}}
\put(12,66){\framebox(4,4){$$}}
\put(36,66){\framebox(4,4){$$}}
\put(60,66){\framebox(4,4){$$}}
%
\put(38,44){\line(0,1){2}}
\put(12,46){\framebox(4,4){$$}}
\put(36,46){\framebox(4,4){$$}}
\put(60,46){\framebox(4,4){$$}}
\put(14,24){\line(0,1){2}}
\put(38,24){\line(0,1){2}}
\put(12,26){\framebox(4,4){$$}}
\put(36,26){\framebox(4,4){$$}}
\put(60,26){\framebox(4,4){$$}}
\put(14,4){\line(0,1){2}}
\put(62,4){\line(0,1){2}}
\put(12,6){\framebox(4,4){$$}}
\put(36,6){\framebox(4,4){$$}}
\put(60,6){\framebox(4,4){$$}}
\put(0,52){\line(-1,0){2}}
\put(24,52){\line(-1,0){2}}
\put(48,52){\line(-1,0){2}}
\put(72,52){\line(-1,0){2}}
\put(-6,50){\framebox(4,4){$$}}
\put(18,50){\framebox(4,4){$$}}
\put(42,50){\framebox(4,4){$$}}
\put(66,50){\framebox(4,4){$$}}
\put(0,32){\line(-1,0){2}}
\put(48,32){\line(-1,0){2}}
\put(-6,30){\framebox(4,4){$$}}
\put(18,30){\framebox(4,4){$$}}
\put(42,30){\framebox(4,4){$$}}
\put(66,30){\framebox(4,4){$$}}
\put(0,12){\line(-1,0){2}}
\put(72,12){\line(-1,0){2}}
\put(-6,10){\framebox(4,4){$$}}
\put(18,10){\framebox(4,4){$$}}
\put(42,10){\framebox(4,4){$$}}
\put(66,10){\framebox(4,4){$$}}
%
 \linethickness{0.65mm}
\put(4,62){\line(1,0){8}}        
\put(16,62){\line(1,0){8}}
\put(28,62){\line(1,0){8}}       
\put(40,62){\line(1,0){8}}
\put(50,54){\line(0,1){6}}
\put(50,50){\line(0,-1){6}}
\put(52,42){\line(1,0){8}}
\put(64,42){\line(1,0){8}}
\put(74,34){\line(0,1){6}}
\put(74,30){\line(0,-1){6}}
\put(28,2){\line(1,0){8}}
\put(40,2){\line(1,0){8}}
\put(26,14){\line(0,1){6}}
\put(26,10){\line(0,-1){6}}
\put(26,34){\line(0,1){6}}
\put(26,30){\line(0,-1){6}}
\put(4,42){\line(1,0){8}}
\put(16,42){\line(1,0){8}}
\put(2,54){\line(0,1){6}}
\put(2,50){\line(0,-1){6}}
\put(50,14){\line(0,1){6}}
\put(50,10){\line(0,-1){6}}
\put(64,22){\line(1,0){8}}
\put(52,22){\line(1,0){8}}
\color{blue}
 \linethickness{0.47mm}
\put(14,64){\line(0,1){2}}
\put(38,64){\line(0,1){2}}
\put(48,52){\line(-1,0){2}}
\put(14,44){\line(0,1){2}}
\put(62,44){\line(0,1){2}}
\put(24,32){\line(-1,0){2}}
\put(72,32){\line(-1,0){2}}
\put(62,24){\line(0,1){2}}
\put(24,12){\line(-1,0){2}}
\put(48,12){\line(-1,0){2}}
\put(38,4){\line(0,1){2}}
\put(0,52){\line(-1,0){2}}
\end{picture}
\vspace{1.3ex}
\caption{\label{fig:2DGridOrig}%
Thick edges show a cycle in the primal NFG of the 2D Ising model, where  
variables attached to the zero-sum indicator factors along the cycle are marked blue.}
\end{figure}
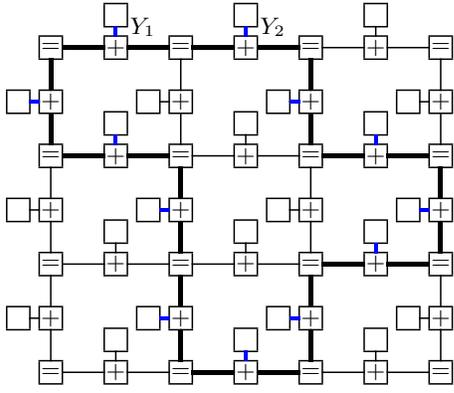

By analogy with Kirchhoff's voltage law, we prove:
\begin{trivlist}
\item{\bf Lemma 1. } Consider a cycle in the primal NFG of the Ising model. If
the variables attached to the zero-sum indicator factors along the cycle are 
$Y_1, Y_2, \ldots$, it holds that 
\begin{IEEEeqnarray}{c}
\label{eqn:cycle}
\sum_\text{$m \in$ Cycle} Y_{m} = 0
\end{IEEEeqnarray}
\end{trivlist}

\begin{trivlist}
\item \emph{Proof. } 
We write each $Y_m$ 
as the addition of its corresponding 
edges $(X_k, X_\ell)$ attached to the zero-sum indicator factors along the cycle (see~(\ref{eqn:XOR})). 
Each variable, say $X_k$, will appear twice in the summation. Thus $\sum_\text{$m \in$ Cycle} Y_{m} = 0$.
\hfill$\blacksquare$
\end{trivlist}

An example of a cycle is shown by thick 
edges in~\Fig{fig:2DGridOrig}, where the variables $Y_1, Y_2, \ldots$ attached 
to the zero-sum indicator factors along the cycle 
are marked blue.

Let us partition $G$ into $G = T \cup \overbar T$, where $T$ is a spanning tree in
the primal NFG. Thus $\Y$ will also be partitioned into \mbox{$\Y_T \cup \Y_{\overbar T}$}.
Examples of such partitionings are shown in~\Fig{fig:2DGridMod}, where spanning trees 
are marked by thick black edges, edges 
attached to the unlabeled boxes and to the zero-sum indicator factors on the branches represent $\Y_T$, and 
edges attached to the unlabeled boxes and to the zero-sum indicator factors on the chords represent $\Y_{\overbar T}$. 

For a given configuration $\y_T$, adding a chord $c \in \overbar T$ to $T$ will create a unique 
cycle called the fundamental 
cycle associated with $c$, which contains exactly one chord that does not appear 
in any other fundamental cycle\footnote{Indeed, the set of all fundamental cycles generates a vector space 
over $\mathbb{F}_2$ with dimensionality $|\overbar T|$; see~\cite[Chapter 2]{Bolob},~\cite[Chapter 14]{Godsil}.}. Furthermore, 
according to Lemma 1, for each $c \in \overbar T$ we can compute $y_c$ as a linear combination of 
$\y_T$. 

\begin{trivlist}
\item{\bf Remark 1. } In the primal NFG, we can freely choose a 
configuration $\y_T$, and 
therefrom deterministically compute each component of $\y_{\overbar T}$ via its 
fundamental cycle. As a result, computing the exact value of $Z$ in the primal NFG requires a sum with 
$|\calX|^{|T|} = |\calX|^{N-1}$ terms.
\end{trivlist}



Accordingly, let
\begin{IEEEeqnarray}{c} \label{eqn:PartG}
\Upsilon(\y) = \prod_{e \in E} \upsilon _{e}(y_e). \label{eqn:PartT}
\end{IEEEeqnarray}



The global probability mass function in the modified primal NFG 
can then be defined as
\begin{IEEEeqnarray}{c}
\label{eqn:GlobDist}
p_{\text{M}}(\y)  \eqdef \frac{\Upsilon(\y)}{Z_\text{M}},
\end{IEEEeqnarray}
where the partition function $Z_\text{M}$ is given by
\begin{IEEEeqnarray}{c}
\label{eqn:ZMdef}
Z_\text{M} = \sum_{\text{valid $\y$}}\Upsilon(\y).
\end{IEEEeqnarray}
%


\begin{trivlist}
\item{\bf Lemma 2. } The partition functions $Z$ 
and $Z_{\text{M}}$ 
are related by
\begin{IEEEeqnarray}{c}
\label{eqn:ScalePrimal}
Z = 2Z_{\text{M}}
\end{IEEEeqnarray}
\end{trivlist}

\begin{trivlist}
\item \emph{Proof. } 
Let $\neg \x$ be the component-wise addition of $\x$ and the all-ones 
vector,  i.e., in $\neg \x$, components of $\x$ that 
are $0$ become $1$, and those that are $1$ become $0$.
There are $|\calX|^{N}$ configurations $\x$ that contribute to $Z$ in~(\ref{eqn:PartFunction}). 
Let us partition $\calX^{N}$ into $\calX_1$ and $\calX_2$, where 
for each $\x \in \calX_1$, we have $\neg \x \in \calX_2$, and vice-versa. 
Note that $|\calX_1| = |\calX_2| = |\calX|^{N-1}$.

There are $|\calX|^{N-1}$ configurations $\y$ with non-zero contributions
to $Z_{\text{M}}$ in~(\ref{eqn:ZMdef}). From one such configuration $\y$, we can compute exactly two corresponding 
configurations $\x$ and $\neg \x$ in the primal NFG (e.g., by setting $x_1 = 0$ and $x_1 = 1$ to solve a 
system of equations for $\x$ and for $\neg \x$). However, due to symmetry in the 
factors~(\ref{eqn:IsingA}), the contribution of $\x$ and of $\neg \x$ to $Z$ is exactly $\Upsilon(\y)$.
\hfill$\blacksquare$
\end{trivlist}

For 2D lattices, Lemmas 1 and 2 have already been observed in~\cite{wu1976}.
However, as was shown, their generalizations to models with arbitrary topology is straightforward.

Similar results can be obtained for the \mbox{$q$-state} Potts model, in which
$\upsilon_{k, \ell}(\cdot)$ is only a function of \mbox{$x_k - x_{\ell}$} (with arithmetic manipulations 
done modulo $q$).
It can be shown that, for this model, $Z = qZ_{\text{M}}$ and 
dependency among the variables is along the cycles 
of a directed NFG.

We next propose a uniform sampling algorithm to compute 
an estimate of $Z_{\text{M}}$, and hence of $Z$ itself. 

\section{Uniform Sampling  in the Primal NFG}
\label{sec:IS}

In uniform sampling, we first draw independent samples $\y^{(1)}_T, \y^{(2)}_T, \ldots$ 
uniformly over $\calX^{|T|}$, i.e., according to
\begin{IEEEeqnarray}{c}
\label{eqn:UP} 
   u_{T}(\y_T) = \frac{1}{|\calX|^{|T|}},
\end{IEEEeqnarray}
and therefrom 
compute $\y^{(1)}_{\overbar T}, \y^{(2)}_{\overbar T}, \ldots$. 
The created samples are
then used in
\begin{IEEEeqnarray}{c}
\label{eqn:EstP} 
   \hat Z_{\text{M}} = \frac{|\calX|^{|T|}}{L}\displaystyle\sum_{\ell = 1}^L\Upsilon(\y^{(\ell)}), 
\end{IEEEeqnarray}
which is an unbiased estimator of $Z_{\text{M}}$, i.e., $\E_{u_T}[\, \hat Z_{\text{M}}\,]  = Z_\text{M}$.

The variance of (\ref{eqn:EstP}) can be computed as
\begin{IEEEeqnarray}{rCl}
\Var[\hat Z_\text{M}] 
 & = & \E\big[\hat Z^2_\text{M}\big] - \E\big[\hat Z_\text{M}\big]^2, \\
 & = & \frac{|\calX|^{2|T|}}{L} \left( \sum_{\y} u_T(\y)\Upsilon(\y)^2 \right) - \frac{Z_\text{M}^2}{L}, \\
 & = & \frac{Z_\text{M}^2}{L}  \left(
        \sum_{\y} 
        \frac{ p_\text{M}(\y)^2}{u_T(\y)} 
        - 1  \right).
\end{IEEEeqnarray}

We thus obtain
\begin{IEEEeqnarray}{c}
\frac{L}{Z^2_{\text{M}}} \V [\hat Z_{\text{M}}] =  \chi^2\big(p_{\text{M}}, u_T\big), \label{eqn:VarBoundP}  
\end{IEEEeqnarray}
where $\chi^2(\cdot, \cdot)$ denotes the chi-square distance, which is non-negative, 
with equality to zero if and only if its two arguments are equal~\cite[Chapter 4]{CS:04}.

Suppose the model is homogeneous (i.e., with constant coupling parameter $J$). In 
the limit $J \to 0$, $\Upsilon(\cdot)$ 
becomes a constant factor (cf.~(\ref{eqn:IsingKernelOrig}), (\ref{eqn:PartG})), therefore we expect
the uniform sampling estimator to perform well when coupling parameters are 
small (i.e., at high temperature). 

Indeed
\begin{IEEEeqnarray}{c}
\lim_{J \to 0}  \chi^2\big(p_{\text{M}}, u_T\big) = 0. \label{eqn:ChiLim}
\end{IEEEeqnarray}

\section{Variance of the Uniform Sampling Algorithm\\
in the Primal 2D Ising Model}
\label{sec:VarIsing}

We analyze the variance of the uniform sampling estimator in the primal domain to 
estimate the partition function of the Ising model on a 2D torus,
with constant coupling parameter $J$ and in the thermodynamic limit (i.e., as $N \to \infty$). 
The choice of the model and the parameters is due to the fact that the
partition function is analytically available from 
Onsager's solution in this case~\cite{Onsager:44},\cite[Chapter 7]{Baxter07}. 
In a 2D torus, it holds that
$|T| = N-1$ and $|\overbar{T}| = N+1$.


From~(\ref{eqn:VarBoundP}), we have
\begin{IEEEeqnarray}{rCl}
\frac{L}{Z_{\text{M}}(J)^2}\Var[\hat Z_{\text{M}}]  & = & \sum_{\text{valid $\y$}}\frac{p_{\text{M}}(\y)^2}{u_T(\y)} - 1,\\
& = & \frac{|\calX|^{|T|}}{Z_{\text{M}}(J)^2} \sum_{\text{valid $\y$}} \Upsilon(\y)^2 -1, \label{eqn:VarPrimalSubs3}\\
& = & |\calX|^{N-1}\frac{Z_{\text{M}}(2J)}{Z_{\text{M}}(J)^2}-1, \label{eqn:VarPrimalSubs4}
\end{IEEEeqnarray}
where $Z_{\text{M}}(J)$ denotes the partition function evaluated at $J$, and the last step is due to the following identity
\begin{equation} \label{eqn:PartFunctionPower}
Z_{\text{M}}(2J) = \sum_{\text{valid $\y$}} \Upsilon(\y)^2.
\end{equation}

Thus, in the thermodynamic limit we obtain
\begin{multline} \label{eqn:VarUnifChi2divPLimit}
\lim_{N \to \infty} \frac{1}{N} \ln \Big(1+\frac{L}{Z_{\text{M}}(J)^2}\Var[\hat Z_{\text{M}}] \Big) = \\ 
\ln(2) + \lim_{N \to \infty}\frac{\ln Z_{\text{M}}(2J)}{N} - \lim_{N \to \infty}\frac{2\ln Z_{\text{M}}(J)}{N}\cdot
\end{multline}

We use the closed-form solution of the partition function to evaluate~(\ref{eqn:VarUnifChi2divPLimit}) numerically as a
function of $J$, which is plotted by the solid black line in~\Fig{fig:Var}. As expected, we observe 
that uniform sampling in the primal domain can provide good estimates of the partition 
function when $J$ is small (i.e., at high temperature), while it is an inefficient estimator 
for larger values of $J$ (i.e., at low temperature).

\section{The Ising Model in the dual Domain}
\label{sec:Dual}


The dual NFG has the same topology as the primal NFG, but  
factors are replaced by the discrete Fourier transform (DFT) of their corresponding 
factors in the primal NFG,
and variables are replaced by their corresponding dual variables, which are 
denoted by the tilde symbol. The partition function of the corresponding dual NFG 
is denoted by $Z_\text{d}$.


According to the normal factor graph duality
theorem~\cite{AY:2011}, $Z_{\text{d}} = \alpha(G)\cdot Z$, where the scale
factor $\alpha(G)$ depends on the topology of $G$ and is given by
\begin{equation}
\label{eqn:scaleF}
\alpha(G) = |\calX|^{|E|-|V|}.
\end{equation} 
The proof of~(\ref{eqn:scaleF}) is given in the Appendix.
For example, for a 2D torus $|E| = 2N$, and therefore $\alpha(G) = |\calX|^{N}$; for 
a chain $|E| = |V|$, and thus $\alpha(G) = 1$.

Notice that from~(\ref{eqn:SizeS}), (\ref{eqn:ScalePrimal}), 
and~(\ref{eqn:scaleF}), we obtain 
\begin{IEEEeqnarray}{rCl}
Z_{\text{d}}/ Z_{\text{M}} & = & \alpha(G)/|\calX|,\\
& = & |\calX|^{|\overbar{T}|}. \label{eqn:scaleMod}
\end{IEEEeqnarray}



From the primal NFG of an Ising model, we can obtain its dual by
replacing each factor~(\ref{eqn:IsingKernelOrig}) by its 1D DFT, each equality indicator 
factor by a zero-sum indicator factor, and each zero-sum indicator factor by an 
equality indicator factor.

\begin{figure}
\begin{picture}(102.5,58.5)(-0.5,0)
 \linethickness{0.2mm}
 \setlength{\unitlength}{0.765mm}
\small
%
\put(32,62){\framebox(4,4){$+$}}
\put(44,62){\framebox(4,4){$=$}}

\put(56,62){\framebox(4,4){$+$}}  
\put(68,62){\framebox(4,4){$=$}}
\put(80,62){\framebox(4,4){$+$}}
\put(36,64){\line(1,0){8}}        
\put(48,64){\line(1,0){8}}
\put(60,64){\line(1,0){8}}      
\put(72,64){\line(1,0){8}}
%
\put(56,55){\framebox(4,4){$=$}}
%
%
\put(10,42){\line(1,0){8}}
\put(22,42){\line(1,0){8}}
\put(32,34){\line(0,1){6}}
\put(32,30){\line(0,-1){6}}
\put(56,34){\line(0,1){6}}
\put(56,30){\line(0,-1){6}}
\put(10,22){\line(1,0){8}}
\put(22,22){\line(1,0){8}}
\put(34,22){\line(1,0){8}}
\put(46,22){\line(1,0){8}}
\put(8,14){\line(0,1){6}}
\put(8,10){\line(0,-1){6}}
\put(56,14){\line(0,1){6}}
\put(56,10){\line(0,-1){6}}
\put(10,2){\line(1,0){8}}
\put(22,2){\line(1,0){8}}
\put(6,40){\framebox(4,4){$+$}}
\put(18,40){\framebox(4,4){$=$}}
\put(30,40){\framebox(4,4){$+$}}
\put(42,40){\framebox(4,4){$=$}}
\put(54,40){\framebox(4,4){$+$}}
%
\put(6,30){\framebox(4,4){$=$}}
\put(30,30){\framebox(4,4){$=$}}
\put(54,30){\framebox(4,4){$=$}}
%
\put(6,20){\framebox(4,4){$+$}}
\put(18,20){\framebox(4,4){$=$}}
\put(30,20){\framebox(4,4){$+$}}
\put(42,20){\framebox(4,4){$=$}}
\put(54,20){\framebox(4,4){$+$}}
%
\put(6,10){\framebox(4,4){$=$}}
\put(30,10){\framebox(4,4){$=$}}
\put(54,10){\framebox(4,4){$=$}}
%
\put(6,0){\framebox(4,4){$+$}}
\put(18,0){\framebox(4,4){$=$}}
\put(30,0){\framebox(4,4){$+$}}
\put(42,0){\framebox(4,4){$=$}}
\put(54,0){\framebox(4,4){$+$}}
%
%
\put(77,8){\line(1,0){8}}
\put(89,8){\line(1,0){8}}
\put(99.325,10.05){\rotatebox{72}{\line(1,0){7.6}}}
\put(103.2,21.5){\rotatebox{72}{\line(1,0){7.5}}}
\put(105.675,32.99){\rotatebox{-36}{\line(-1,0){7.67}}}
\put(95.2,40.2){\rotatebox{-36}{\line(-1,0){7.71}}}
\put(69.7,30.825){\line(1,0){15.1}}
\put(104.4,30.825){\line(-1,0){15.2}}
\put(73,6){\framebox(4,4){$+$}}
\put(85,6){\framebox(4,4){$=$}}
\put(97,6){\framebox(4,4){$+$}}
%
\put(100.6,17.4){\framebox(4,4){$=$}}
\put(104.416,28.825){\framebox(4,4){$+$}}
%
\put(95.5,36.9){\framebox(4,4){$=$}}
%
\put(74.5,36.9){\framebox(4,4){$=$}}
%
\put(69.6,17.4){\framebox(4,4){$=$}}
\put(65.684,28.825){\framebox(4,4){$+$}}
\put(85,43.932){\framebox(4,4){$+$}}
%
\put(85,28.825){\framebox(4,4){$=$}}
%
\put(81.9,17.4){\framebox(4,4){$=$}}
%
\put(88.2,17.4){\framebox(4,4){$=$}}
%
\put(78,24.8){\framebox(4,4){$=$}}
%
\put(92,24.8){\framebox(4,4){$=$}}
%
\put(46,66){\line(0,1){2}}
\put(44,68){\framebox(4,4){}}
\put(70,66){\line(0,1){2}}
\put(68,68){\framebox(4,4){}}
\put(58,55){\line(0,-1){2}}
\put(56,49){\framebox(4,4){}}
%
\put(20,44){\line(0,1){2}}
\put(44,44){\line(0,1){2}}
\put(18,46){\framebox(4,4){}}
\put(42,46){\framebox(4,4){}}
\put(6,32){\line(-1,0){2}}
\put(30,32){\line(-1,0){2}}
\put(54,32){\line(-1,0){2}}
\put(0,30){\framebox(4,4){}}
\put(24,30){\framebox(4,4){}}
\put(48,30){\framebox(4,4){}}
\put(20,24){\line(0,1){2}}
\put(44,24){\line(0,1){2}}
\put(18,26){\framebox(4,4){}}
\put(42,26){\framebox(4,4){}}
\put(6,12){\line(-1,0){2}}
\put(30,12){\line(-1,0){2}}
\put(54,12){\line(-1,0){2}}
\put(0,10){\framebox(4,4){}}
\put(24,10){\framebox(4,4){}}
\put(48,10){\framebox(4,4){}}
\put(20,4){\line(0,1){2}}
\put(44,4){\line(0,1){2}}
\put(18,6){\framebox(4,4){}}
\put(42,6){\framebox(4,4){}}
%
%
\put(87,6){\line(0,-1){2}}
\put(85,0){\framebox(4,4){}}
\put(104.6,19.4){\line(1,0){2}}
\put(106.6,17.4){\framebox(4,4){}}
\put(99.5,38.9){\line(1,0){2.2}}
\put(101.7,36.9){\framebox(4,4){}}
\put(69.6,19.4){\line(-1,0){2}}
\put(63.6,17.4){\framebox(4,4){}}
\put(87.02,32.825){\line(0,1){2}}
\put(85.3,34.825){\framebox(3.71,3.71){}}
\put(79.9,19.4){\line(1,0){2}}
\put(75.9,17.4){\framebox(4,4){}}
\put(92.2,19.4){\line(1,0){2}}
\put(94.2,17.4){\framebox(4,4){}}
\put(78,26.8){\line(-1,0){2}}
\put(72,24.8){\framebox(4,4){}}
\put(96,26.8){\line(1,0){2}}
\put(98,24.8){\framebox(4,4){}}
\put(72.5,38.9){\line(1,0){2}}
\put(68.5,36.9){\framebox(4,4){}}
\put(24.75,44.0){\pos{bc}{$\tilde Y_1$}}
\color{blue}
\linethickness{0.48mm}
%
 %
 \put(34,57){\line(0,1){5}}
\put(34,57){\line(1,0){22}}
\put(82,57){\line(0,1){5}}
\put(60,57){\line(1,0){22}}
\put(34,42){\line(1,0){8}}
\put(46,42){\line(1,0){8}}
\put(8,34){\line(0,1){6}}
\put(8,30){\line(0,-1){6}}
\put(32,14){\line(0,1){6}}
\put(32,10){\line(0,-1){6}}
\put(34,2){\line(1,0){8}}
\put(46,2){\line(1,0){8}}
%
%
\put(68.6,32.925){\rotatebox{37}{\line(1,0){7.25}}}
\put(78.5,40.4){\rotatebox{37}{\line(1,0){7.8}}}
\put(74.675,10.05){\rotatebox{-72}{\line(-1,0){7.6}}}
\put(70,21.5){\rotatebox{-72}{\line(-1,0){7.5}}}
\put(97.675,10.05){\rotatebox{-35}{\line(-1,0){14.3}}}
\put(69.8,29.55){\rotatebox{-36}{\line(1,0){14.7}}}
\put(76.3,10.05){\rotatebox{35}{\line(1,0){14.25}}}
\put(104.2,29.55){\rotatebox{36}{\line(-1,0){14.75}}}
\put(75.1,10.05){\rotatebox{72}{\line(1,0){15.4}}}
\put(86.1,43.9){\rotatebox{72}{\line(-1,0){15.7}}}
\put(99,10.05){\rotatebox{-72}{\line(-1,0){15.4}}}
\put(87.95,43.7){\rotatebox{-72}{\line(1,0){15.7}}}
%
%
\end{picture}
\vspace{1.0ex}
\caption{\label{fig:2DGridD}%
Dual NFGs of the Ising models in~\Fig{fig:2DGridMod}.
The unlabeled boxes represent~(\ref{eqn:IsingKernelD}), boxes containing 
$``+"$ symbols are as in~(\ref{eqn:XOR}), and boxes containing $``="$ symbols 
are given by~(\ref{eqn:equality}). In each dual  
NFG, the chords are marked by thick blue edges.}
\end{figure}
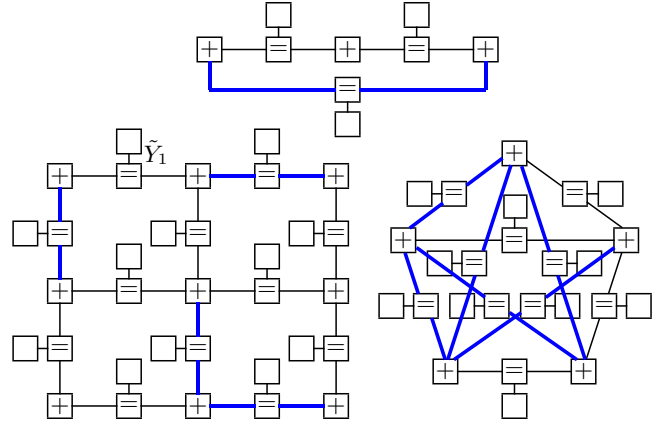

The dual NFGs of the Ising models in~\Fig{fig:2DGridMod} are shown 
in~\Fig{fig:2DGridD}, where the 
unlabeled boxes represent factors as
\begin{equation} 
\label{eqn:IsingKernelD}
\gamma_{e}(\tilde y_e) = \left\{ \begin{array}{ll}
      2\cosh{J_e}, & \text{if $\tilde y_e = 0$} \\
      2\sinh{J_e}, & \text{if $\tilde y_e = 1$,}
  \end{array} \right.
\end{equation}
boxes labeled ``$+$'' are zero-sum indicator factors as in~(\ref{eqn:XOR}), and
boxes containing $``="$ 
symbols are equality indicator factors given by~(\ref{eqn:equality}). 
For more details on constructing the dual NFG of the Ising model, see~\cite{MoLo:ISIT2013, AY:2014, MeMo:2014a, Mo:IZS2016}.

By analogy with Kirchhoff's current 
law, we prove:
\begin{trivlist}
\item{\bf Lemma 3. } Consider a cutset in the dual NFG of the Ising model. If 
the variables attached to the equality indicator factors in the cutset 
are $\tilde Y_1, \tilde Y_2, \ldots$, it holds that
\begin{IEEEeqnarray}{c}
\label{eqn:cut}
\sum_\text{$m \in$ Cutset} \tilde Y_{m} = 0
\end{IEEEeqnarray}
\end{trivlist}

\begin{trivlist}
\item \emph{Proof. } 
A cutset partitions $G$ into $G_1\cup \,G_2$. In $G_1$ (or in $G_2$), suppose we 
write down the equations associated with all the zero-sum indicator factors. 
But the sum over all these equations in $G_1$ (or in $G_2$) 
is equal to zero, because each variable, say $\tilde Y_k$, appears twice in the 
summation. Furthermore, in $G$, the same sums are equal to $\sum_\text{$m \in$ Cutset} \tilde Y_{m}$. 
\hfill$\blacksquare$
\end{trivlist}

\begin{figure}[t]
\setlength{\unitlength}{0.73mm}
\centering
\begin{picture}(77,73)(0,0)
 \linethickness{0.2mm}
\small
\put(0,60){\framebox(4,4){$+$}}
\put(12,60){\framebox(4,4){$=$}}  
\put(24,60){\framebox(4,4){$+$}}
\put(36,60){\framebox(4,4){$=$}}
\put(48,60){\framebox(4,4){$+$}}
\put(52,62){\line(1,0){8}}       
\put(60,60){\framebox(4,4){$=$}} 
\put(64,62){\line(1,0){8}}
\put(72,60){\framebox(4,4){$+$}}
%
\put(0,50){\framebox(4,4){$=$}}
\put(26,54){\line(0,1){6}}
\put(24,50){\framebox(4,4){$=$}}
\put(26,50){\line(0,-1){6}}
\put(50,54){\line(0,1){6}}
\put(48,50){\framebox(4,4){$=$}}
\put(50,50){\line(0,-1){6}}
\put(74,54){\line(0,1){6}}
\put(72,50){\framebox(4,4){$=$}}
\put(74,50){\line(0,-1){6}}
\put(0,40){\framebox(4,4){$+$}}
\put(12,40){\framebox(4,4){$=$}}
\put(24,40){\framebox(4,4){$+$}}
\put(36,40){\framebox(4,4){$=$}}
\put(48,40){\framebox(4,4){$+$}}
\put(60,40){\framebox(4,4){$=$}}
\put(72,40){\framebox(4,4){$+$}}
%
\put(2,34){\line(0,1){6}}
\put(0,30){\framebox(4,4){$=$}}
\put(2,30){\line(0,-1){6}}
\put(24,30){\framebox(4,4){$=$}}
\put(48,30){\framebox(4,4){$=$}}
\put(72,30){\framebox(4,4){$=$}}
%
\put(0,20){\framebox(4,4){$+$}}
\put(12,20){\framebox(4,4){$=$}}
\put(24,20){\framebox(4,4){$+$}}
\put(36,20){\framebox(4,4){$=$}}
\put(48,20){\framebox(4,4){$+$}}
\put(60,20){\framebox(4,4){$=$}}
\put(72,20){\framebox(4,4){$+$}}
%
\put(2,14){\line(0,1){6}}
\put(0,10){\framebox(4,4){$=$}}
\put(2,10){\line(0,-1){6}}
\put(24,10){\framebox(4,4){$=$}}
\put(48,10){\framebox(4,4){$=$}}
\put(74,14){\line(0,1){6}}
\put(72,10){\framebox(4,4){$+$}}
\put(74,10){\line(0,-1){6}}
\put(0,0){\framebox(4,4){$+$}}
\put(4,2){\line(1,0){8}}
\put(12,0){\framebox(4,4){$=$}}
\put(16,2){\line(1,0){8}}
\put(24,0){\framebox(4,4){$+$}}
\put(36,0){\framebox(4,4){$=$}}
\put(48,0){\framebox(4,4){$+$}}
\put(60,0){\framebox(4,4){$=$}}
\put(72,0){\framebox(4,4){$+$}}
 
\put(14,64){\line(0,1){2}}
\put(62,64){\line(0,1){2}}
\put(12,66){\framebox(4,4){$$}}
\put(36,66){\framebox(4,4){$$}}
\put(60,66){\framebox(4,4){$$}}
%
\put(12,46){\framebox(4,4){$$}}
\put(36,46){\framebox(4,4){$$}}
\put(60,46){\framebox(4,4){$$}}
%
\put(12,26){\framebox(4,4){$$}}
\put(36,26){\framebox(4,4){$$}}
\put(60,26){\framebox(4,4){$$}}
\put(14,4){\line(0,1){2}}
\put(38,4){\line(0,1){2}}
\put(12,6){\framebox(4,4){$$}}
\put(36,6){\framebox(4,4){$$}}
\put(60,6){\framebox(4,4){$$}}
%
\put(24,52){\line(-1,0){2}}
\put(72,52){\line(-1,0){2}}
\put(-6,50){\framebox(4,4){$$}}
\put(18,50){\framebox(4,4){$$}}
\put(42,50){\framebox(4,4){$$}}
\put(66,50){\framebox(4,4){$$}}
\put(0,32){\line(-1,0){2}}
\put(-6,30){\framebox(4,4){$$}}
\put(18,30){\framebox(4,4){$$}}
\put(42,30){\framebox(4,4){$$}}
\put(66,30){\framebox(4,4){$$}}
\put(0,12){\line(-1,0){2}}
\put(72,12){\line(-1,0){2}}
\put(-6,10){\framebox(4,4){$$}}
\put(18,10){\framebox(4,4){$$}}
\put(42,10){\framebox(4,4){$$}}
\put(66,10){\framebox(4,4){$$}}
\put(4,62){\line(1,0){8}}        
\put(16,62){\line(1,0){8}}
\put(50,54){\line(0,1){6}}
\put(50,50){\line(0,-1){6}}
\put(74,34){\line(0,1){6}}
\put(74,30){\line(0,-1){6}}
\put(28,2){\line(1,0){8}}
\put(40,2){\line(1,0){8}}
\put(26,14){\line(0,1){6}}
\put(26,10){\line(0,-1){6}}
\put(26,34){\line(0,1){6}}
\put(26,30){\line(0,-1){6}}
\put(4,42){\line(1,0){8}}
\put(16,42){\line(1,0){8}}
\put(2,54){\line(0,1){6}}
\put(2,50){\line(0,-1){6}}
\put(50,14){\line(0,1){6}}
\put(50,10){\line(0,-1){6}}
%
\put(48,52){\line(-1,0){2}}
\put(14,44){\line(0,1){2}}
\put(24,32){\line(-1,0){2}}
\put(72,32){\line(-1,0){2}}
\put(24,12){\line(-1,0){2}}
\put(48,12){\line(-1,0){2}}
\put(0,52){\line(-1,0){2}}
\put(52,42){\line(1,0){8}}
\put(64,42){\line(1,0){8}}
\put(4,22){\line(1,0){8}}
\put(16,22){\line(1,0){8}}
\put(28,22){\line(1,0){8}}
\put(40,22){\line(1,0){8}}
\put(62,44){\line(0,1){2}}
\put(14,24){\line(0,1){2}}
\put(38,24){\line(0,1){2}}
 \linethickness{0.65mm}
%
\put(28,62){\line(1,0){8}}       
\put(40,62){\line(1,0){8}}
\put(28,42){\line(1,0){8}}
\put(40,42){\line(1,0){8}}
\put(52,2){\line(1,0){8}}
\put(64,2){\line(1,0){8}}
\put(50,34){\line(0,1){6}}
\put(50,30){\line(0,-1){6}}
\put(64,22){\line(1,0){8}}
\put(52,22){\line(1,0){8}}
\color{blue}
 \linethickness{0.47mm}
\put(48,32){\line(-1,0){2}}
\put(62,24){\line(0,1){2}}
\put(62,4){\line(0,1){2}}
\put(38,64){\line(0,1){2}}
\put(38,44){\line(0,1){2}}
\end{picture}
\vspace{1.0ex}
\caption{\label{fig:2DGridDual}%
Thick edges show a cutset in the dual NFG of the 2D Ising model, where  
variables on the cutset are marked blue.}
\end{figure}
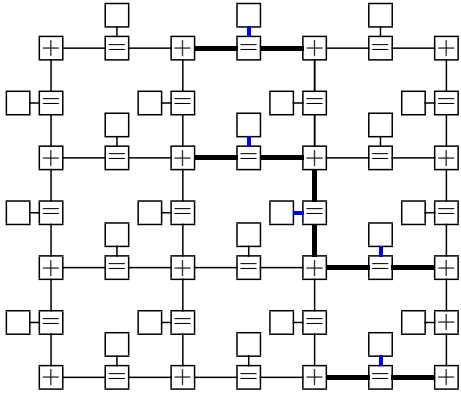

An example of a cutset is shown by thick 
edges in~\Fig{fig:2DGridDual}, where the variables $\tilde Y_1, \tilde Y_2, \ldots$ attached to the 
equality indicator factors in the cutset are marked blue.

Again, we partition $G$ into $G = T \cup \overbar T$, where $T$ is a spanning tree in
the dual NFG. As a result, $\tilde \Y = \tilde \Y_T \cup \tilde \Y_{\overbar T}$.
\Fig{fig:2DGridD} shows examples of such partitionings, where cospanning trees 
are marked by thick blue edges, edges 
attached to the unlabeled boxes and to the equality indicator factors on the branches represent $\tilde \Y_T$, and 
edges attached to the unlabeled boxes and to the equality indicator factors on the chords 
represent $\tilde \Y_{\overbar T}$. Although $T$ is always cycle-free, $\overbar T$ may contain 
cycles (see~\Fig{fig:2DGridD}--right). 

Removing a branch $b \in T$ partitions $T=T_1 \cup T_2$. The edges that connect $T_1$ and 
$T_2$ form a unique cutset in $G$ -- called the fundamental 
cutset belonging to $b$. Each fundamental cutset has exactly one branch of $T$ that does not appear 
in any other fundamental cutset\footnote{In the dual domain, the set of all fundamental cutsets generates a vector space 
over $\mathbb{F}_2$ with dimensionality $|T|$; see~\cite[Chapter 2]{Bolob},~\cite[Chapter 14]{Godsil}.}. Moreover, 
according to Lemma 3, for each $b \in T$ we can compute $\tilde y_b$ as a linear combination of 
$\tilde \y_{\overbar T}$. 

\begin{trivlist}
\item{\bf Remark 2. } In the dual NFG, we can freely choose a configuration 
$\tilde \y_{\overbar T}$, and 
therefrom deterministically compute each component of $\tilde \y_{T}$ via its 
fundamental cutset. In the dual NFG, computing the exact value of $Z_\text{d}$ (and thus the 
exact value of $Z_\text{M}$) 
requires a 
sum with $|\calX|^{|\overbar T|}$ terms. In particular, computing $Z_\text{d}$ of a chain requires 
a sum with $|\calX|$ terms.
\end{trivlist}

Let
\begin{IEEEeqnarray}{c}
\Gamma(\tilde \y) =  \prod_{e \in E} \gamma _{e}(\tilde y_e). \label{eqn:PartTD}
\end{IEEEeqnarray}

Suppose the model is ``ferromagnetic" (i.e., \mbox{$J_{k,\ell} > 0$} for all 
$(k, \ell) \in E$), thus $\Gamma(\cdot)$ is 
non-negative. We then define the following global probability mass function in the dual NFG
\begin{IEEEeqnarray}{c}
\label{eqn:GlobDistD}
p_{\text{d}}(\tilde \y) \eqdef \frac{\Gamma(\tilde \y)}{Z_\text{d}},
\end{IEEEeqnarray}
where 
\begin{IEEEeqnarray}{c}
\label{eqn:GlobDistD}
Z_\text{d} = \sum_{\text{valid $\tilde \y$}}\Gamma(\tilde \y).
\end{IEEEeqnarray}

Next, we propose a uniform sampling algorithm in the dual NFG to estimate $Z_{\text{d}}$.

%
%
%
%

\section{Uniform Sampling in the Dual NFG}
\label{sec:ISD}

In uniform sampling in the dual domain, samples $\y^{(1)}_{\overbar T}, \y^{(2)}_{\overbar T}, \ldots$
are drawn independently according to
\begin{IEEEeqnarray}{c}
\label{eqn:UPD} 
   u_{\overbar{T}}(\tilde \y_{\overbar{T}}) = \frac{1}{|\calX|^{|\overbar{T}|}},
\end{IEEEeqnarray}
and completed to valid configurations $\y^{(1)}, \y^{(2)}, \ldots$. Then $L$ created samples are used in the 
following estimator
\begin{IEEEeqnarray}{c}
\label{eqn:EstimDual} 
\hat Z_{\text{d}}  =  \frac{|\calX|^{|\overbar{T}|}}{L}\displaystyle\sum_{\ell = 1}^L\Gamma(\tilde \y^{(\ell)}),
\end{IEEEeqnarray}
which is unbiased, that is, $\E_{u_{\overbar T}}[\, \hat Z_{\text{d}}\,] = Z_{\text{d}}$ (see~\cite{Mo:IZS2016}).

The variance of~(\ref{eqn:EstimDual}) is given by
\begin{IEEEeqnarray}{c}
\frac{L}{Z^2_{\text{d}}} \V [\hat Z_{\text{d}}] = \chi^2\big(p_{\text{d}}, u_{\overbar T}\big). \label{eqn:VarBoundD}  
\end{IEEEeqnarray}

In the low temperature limit $p_{\text{d}}$ becomes uniform over the
valid configurations (cf.~(\ref{eqn:IsingKernelD}),~(\ref{eqn:PartTD})). The estimator is thus expected to perform well in the low-temperature regime (i.e., for large $J$).
Indeed
\begin{IEEEeqnarray}{c}
\label{eqn:ChiD}
\lim_{J \to \infty} \chi^2\big(p_{\text{d}}, u_{\overbar T}\big) =  0.  
\end{IEEEeqnarray}

\section{Variance of the Uniform Sampling Algorithm
in the Dual 2D Ising Model}
\label{sec:VarIsingD}

In the dual domain,  we provide upper 
and lower bounds on the variance of the estimator. The derived bounds are not necessarily tight
for all values of $J$; however, they are good enough to illustrate the opposite behavior 
of~(\ref{eqn:EstP}) and (\ref{eqn:EstimDual}). 

From~(\ref{eqn:VarBoundD}), we have
\begin{IEEEeqnarray}{rCl}
 \frac{L}{Z_\text{d}(J)^2}\Var[\hat Z_\text{d}] & = & \sum_{\text{valid $\tilde \y$}} \frac{p_\text{d}(\tilde\y)^2}{u_{\overbar{T}}(\tilde \y)} -1, \\
& = & \frac{|\calX|^{|\overbar{T}|}}{Z_\text{d}(J)^2} \sum_{\text{valid $\tilde \y$}} \Gamma(\tilde\y)^2 -1, \label{eqn:VarDualSubs2}\\
& = & \frac{2^{N+1}}{Z_\text{d}(J)^2} S_\text{d} -1, \label{eqn:VarDualSubs3}
\end{IEEEeqnarray}
where $S_\text{d} \eqdef \sum_{\text{valid $\tilde \y$}} \Gamma(\tilde\y)^2$.

From~(\ref{eqn:scaleF}) and~(\ref{eqn:ScalePrimal}), we obtain $Z_\text{d} = 2^{N+1}Z_{\text{M}}$, therefore 
\begin{IEEEeqnarray}{c}
\frac{L}{Z_\text{d}(J)^2} \Var[\hat Z^{\text{Uni}}_\text{d}] = 
\frac{2^{-N-1}}{Z_{\text{M}}(J)^2}S_\text{d} -1. \label{eqn:VarDualSubs}
\end{IEEEeqnarray}

\begin{figure}[t]
\centering
\begin{tikzpicture}
\begin{axis}[
			legend style={at = {(0.99,0.71)} ,font=\tiny},		
			height = 46.0ex,
			width = 53.0ex,
			grid = major,
			tick pos=left, 
			xminorticks = false,	
		    yminorticks = false,	
		    y tick label style={
        /pgf/number format/.cd,
            fixed,
        /tikz/.cd
    		}, 				
			ytick={0,  0.2, 0.4, 0.6, 0.8},
			xtick={0.0, 0.5, 1.0, 1.5, 2.0, 2.5, 3.0},
		xlabel= $J$ ={font=\normalsize},
			xmin = 0.0,
			xmax = 2.5,
			ymin = 0.0,
			ymax = 0.8,
			ylabel = $\lim_{N \to \infty}\frac{1}{N} \ln\big(1+ \frac{L}{Z^2}\V \hat Z \big)$ = {font=\normalsize},
			yticklabel style = {font=\tiny,yshift=-0.25ex},
            xticklabel style = {font=\tiny,xshift=-0.1ex}			
			]
\pgfplotstableread{./VarUP.txt}\mydataone
\pgfplotstableread{./VarUDLower.txt}\mydatathree
\pgfplotstableread{./VarUDUpper.txt}\mydatatwo
\pgfplotstableread{./Criticality.txt}\mydatafour
		\addplot [
		 line width = 0.24mm,
		 smooth,
		 color = black,
		]		
		 table[y = Z] from \mydataone;

		 \addplot [
 		 color = blue,
		 line width = 0.1mm,
 		 ]
 		  table[y = Z] from \mydatafour;	 
		 
 		 \addplot [
		 line width = 0.25mm,
		 smooth,
 		 color = blue,
		 densely dotted,
 		 ]
 		  table[y = Z] from \mydatatwo;	 

		 \addplot [
		 line width = 0.25mm,
		 smooth,
 		 color = red,
	     dashed,
 		 ]
 		  table[y = Z] from \mydatathree;

 		 \legend{Primal domain exact result~(\ref{eqn:VarUnifChi2divPLimit}), Dual domain upper bound~(\ref{eqn:VarUnifChi2divDUpper2}), Dual domain upper bound~(\ref{eqn:VarUnifChi2divDUpper}), Dual domain lower bound~(\ref{eqn:VarUnifChi2divDLower})};	  	


\end{axis}
\end{tikzpicture}
\caption{\label{fig:Var}%
Behavior of the variance of the uniform sampling estimator in the primal and in the dual NFGs as a function of 
the coupling parameter $J$, for a homogeneous
2D Ising model in the thermodynamic limit. 
The solid black line shows~(\ref{eqn:VarUnifChi2divPLimit}), the solid blue line shows the upper 
bound in~(\ref{eqn:VarUnifChi2divDUpper2}), the dotted blue line shows the upper 
bound in~(\ref{eqn:VarUnifChi2divDUpper}), and the dashed red line shows the lower 
bound in~(\ref{eqn:VarUnifChi2divDLower}).}
\end{figure}
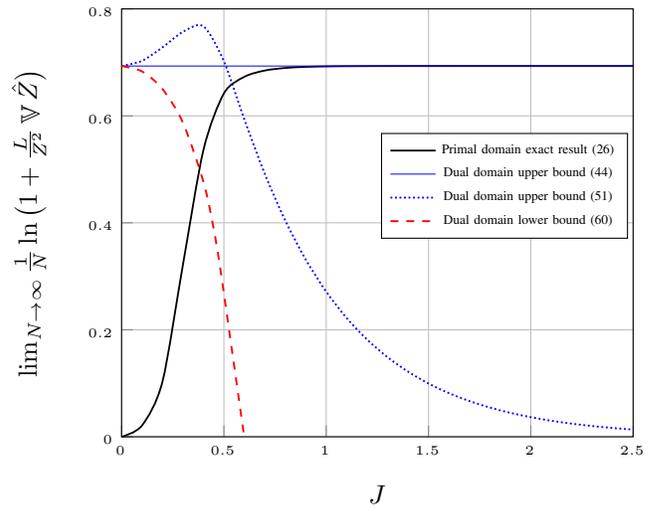

In the rest of this section, we will derive upper and lower bounds on $S_\text{d}$. We first 
apply the obvious inequality 
\begin{equation} \label{eqn:SimpleDUpper}
S_\text{d} \le Z_\text{d}(J)^2
\end{equation}
in~(\ref{eqn:VarDualSubs3}) to obtain
\begin{equation} \label{eqn:VarUnifChi2divDUpper2}
\lim_{N \to \infty} \frac{1}{N}\ln 
\Big(1+ \frac{L}{Z_\text{d}(J)^2}\Var[\hat Z^{\text{Uni}}_\text{d}]\Big) \le \ln(2),
\end{equation}
which is plotted by the solid blue line in~\Fig{fig:Var}.

We next note that $S_\text{d}$ is the partition function of a
dual NFG (as in~\Fig{fig:2DGridD}--left) with factors given by
\begin{equation} 
\label{eqn:IsingDualAddPower}
\rho(\tilde{y}_e) = \left\{ \begin{array}{ll}
      4\cosh(J)^2, & \text{if $\tilde{y}_e = 0$} \\
      4\sinh(J)^2, & \text{if $\tilde{y}_e = 1$.}
  \end{array} \right.
\end{equation}


Thus
\begin{IEEEeqnarray}{rCl}  
S_\text{d} &=& \sum_{\text{valid $\tilde\y$}} \prod_{e \in E} \rho(\tilde y_e), \\
& \le & \big(4\cosh(J)^2\big)^{|T|}\sum_{\tilde\y_{\overbar{T}}} \prod_{e\in \overbar{T}} \rho(\tilde y_e), \\
& = & \big(2\cosh(J)\big)^{2(N-1)}S_{\overbar{T}}. \label{eqn:DualPartitionSumGammaPower}
\end{IEEEeqnarray}

Here, $S_{\overbar{T}}$ is the partition function of a subgraph in the dual NFG induced 
by $\overbar{T}$, which can be
computed exactly as
\begin{IEEEeqnarray}{rCl} 
S_{\overbar{T}} & = & \big(\rho(0) + \rho(1)\big)^{|\overbar{T}|}, \\
        &=& \big(4\cosh(2J)\big)^{N+1}. \label{eqn:ZFisPower}
\end{IEEEeqnarray}

Combining~(\ref{eqn:VarDualSubs}),~(\ref{eqn:DualPartitionSumGammaPower}), and~(\ref{eqn:ZFisPower}) yields 
\begin{multline} \label{eqn:VarUnifChi2divDUpper}
\lim_{N \to \infty} \frac{1}{N}\ln \Big(1+\frac{L}{Z_\text{d}(J)^2}\Var[\hat Z_\text{d}]\Big) \le 3\ln(2) + \\ 
 \ln\big(\cosh(2J)\cdot\cosh(J)^2\big) - \lim_{N \to \infty}\frac{2\ln Z_{\text{M}}(J)}{N},
\end{multline}
which is plotted by the dotted blue line in~\Fig{fig:Var}.

To obtain the lower bound, we consider the corresponding primal NFG (as 
in Fig.~\ref{fig:2DGridMod}--left)
with factors as in
\begin{equation} 
\label{eqn:IsingPrimalAddPower}
\kappa(y_e) = \left\{ \begin{array}{ll}
      2\cosh(2J), & \text{if $y_e = 0$} \\
      2, & \text{if $y_e = 1$.}
  \end{array} \right.
\end{equation}

Notice that~(\ref{eqn:IsingPrimalAddPower}) is indeed the inverse Fourier transform
of~(\ref{eqn:IsingDualAddPower}).
We denote the partition function of this primal NFG by $S_\text{M}$, where according to the NFG duality
theorem
\begin{equation}
S_\text{d}= 2^{N+1}S_{\text{M}}, \label{eqn:ScaleSSd}
\end{equation}
see~(\ref{eqn:scaleMod}). Hence
\begin{IEEEeqnarray}{rCl}
S_{\text{M}} & = & \sum_{\text{valid $\y$}}  \prod_{e \in E} \kappa(y_e), \\
  & \ge & 2^{|\overbar{T}|}\sum_{\text{$\y_T$}}  \prod_{e \in T} \kappa(y_e), \label{eqn:ZFisPrimalPower}\\
& = & 2^{N+1}S_T, \label{eqn:primalSubTreeBound}
\end{IEEEeqnarray}
where $S_T$ denotes the partition function of a subgraph in the primal NFG 
induced by $T$ (i.e., a spanning tree), which again can be computed 
exactly as
\begin{IEEEeqnarray}{rCl}  
S_T &=& \big(\kappa(0) + \kappa(1)\big)^{|T|}, \\
&=& \big(2\cosh(J)\big)^{2N-2}, \label{eqn:STExact}
\end{IEEEeqnarray}
see~\cite[Chapter 2]{Baxter07},~\cite[Section~III]{MoLo:ISIT2013}.

From~(\ref{eqn:ScaleSSd}),~(\ref{eqn:primalSubTreeBound}), and~(\ref{eqn:STExact}) we obtain
\begin{equation}
S_\text{d} \ge 2^{4N}\cosh(J)^{2(N-1)}.
      \label{eqn:ZFisPrimalPower}
\end{equation}

Combining~(\ref{eqn:VarDualSubs}) and (\ref{eqn:ZFisPrimalPower}) gives the following 
lower bound
\begin{multline} \label{eqn:VarUnifChi2divDLower}
\lim_{N \to \infty} \frac{1}{N}\ln \Big(1+ \frac{L}{Z_\text{d}(J)^2}\Var[\hat Z_\text{d}]\Big) \ge \\ 
3\ln2  + 2\ln\big(\cosh(J)\big) - \lim_{N \to \infty}\frac{2\ln Z_{\text{M}}(J)}{N},
\end{multline}
which is shown by the dashed red line in~\Fig{fig:Var}.

From~\Fig{fig:Var}, we observe that uniform sampling in the 
dual domain is inefficient for small values of $J$; however,  compared to uniform sampling 
in the primal domain,
it can provide more reliable estimates of the partition function when $J$ is large. 
Alos, recall from Section~\ref{sec:ISD} that~(\ref{eqn:VarBoundD}) vanishes in the 
low-temperature limit (i.e., as $J \to \infty$).

Both estimators seem to be inefficient in the mid-temperature regime and near 
\emph{criticality}, which for this model is located 
at $J_\text{c} = \frac{1}{2}\ln(1+\sqrt{2}) \approx 0.44$ (see~\cite[Chapter 6]{Baxter07}).

\section{Conclusion}

We analyzed some properties of the Ising model (in the primal and in the dual domains)
in the context of algebraic graph theory. We showed that, in the primal domain, variables
can be freely chosen on a spanning tree, and the remaining variables can be computed 
via their fundamental cycles, whereas in the dual domain, we can choose the variables arbitrarily
on a cospanning tree, and compute the remaining variables via their fundamental 
cutsets. In each domain, a uniform sampling algorithm was proposed to estimate the 
partition function, and its opposite behavior was illustrated for the homogeneous Ising 
model on a 2D torus.

\section*{Appendix\\Details of the scale factor}
\label{appsec:DualDetails}


For completeness, 
we prove that the scale factor between the partition function $Z$ of an NFG and 
the partition function $Z_\text{d}$ of the corresponding dual NFG is
\begin{equation} \label{eqn:AppZdcZ}
\alpha(G) = |\calX|^{|E|-|V|},
\end{equation}
where $\alpha(G) = Z_{\text{d}}/Z$, which depends on the topology of $G$. 

We will use the following concepts: 
a box is a collection of factors as illustrated by the dashed lines in \Fig{fig:Boxes}, and 
the \emph{exterior function} of such a box 
is the product of all factors inside the box, summed over all 
variables inside the box~\cite{AY:2011}. 
For example, the exterior function of the inner dashed box 
in \Fig{fig:Boxes} is given by
\begin{equation} \label{eqn:AppBoxesExg}
g(x_1,x_3,x_5) = \sum_{x_2} f_1(x_1,x_2,x_5) f_2(x_2,x_3),
\end{equation}
and the exterior function of the outer dashed box in \Fig{fig:Boxes}
is given by
\begin{equation} \label{eqn:AppBoxesExZ}
Z = \sum_{x_1,\ldots,x_5} f_1(x_1,x_2,x_5) f_2(x_2,x_3) f_3(x_3,x_4,x_5).
\end{equation}

\begin{figure}[t!!]
\setlength{\unitlength}{0.890mm}
\centering
\begin{picture}(80,32.5)(-5,-10)
\put(-5,-10){\dashbox(80,30){}}
\put(0,2.5){\line(1,0){12.5}}    \put(2.5,3.5){\pos{cb}{$X_1$}}
\put(12.5,0){\framebox(5,5){}}   \put(15,-1.25){\pos{ct}{$f_1$}}
 \put(15,15){\line(0,-1){10}}
 \put(15,15){\line(1,0){40}}     \put(45,14){\pos{ct}{$X_5$}}
 \put(55,15){\line(0,-1){10}}
\put(17.5,2.5){\line(1,0){10}}   \put(22.5,3.5){\pos{cb}{$X_2$}}
\put(27.5,0){\framebox(5,5){}}   \put(30,-1.25){\pos{ct}{$f_2$}}
\put(32.5,2.5){\line(1,0){20}}   \put(45,3.5){\pos{cb}{$X_3$}}
\put(52.5,0){\framebox(5,5){}}   \put(55,-1.25){\pos{ct}{$f_3$}}
\put(57.5,2.5){\line(1,0){12.5}}  \put(65,3.5){\pos{cb}{$X_4$}}
\put(7.5,-6){\dashbox(30,16){}}  \put(38.5,-6){\pos{bl}{$g$}}
\end{picture}
\vspace{2\unitlength}
\caption{\label{fig:Boxes}%
Boxes in an NFG.}
\setlength{\unitlength}{0.89mm}
\centering
\begin{picture}(50,34)(0,-10)
\put(0,2.5){\line(1,0){12.5}}    \put(5,3.5){\pos{cb}{$X_1$}}
\put(12.5,0){\framebox(5,5){}}   \put(15,-1.25){\pos{ct}{$g$}}
 \put(15,15){\line(0,-1){10}}
 \put(15,15){\line(1,0){20}}     \put(25,14){\pos{ct}{$X_5$}}
 \put(35,15){\line(0,-1){10}}
\put(17.5,2.5){\line(1,0){15}}    \put(25,3.5){\pos{cb}{$X_3$}}
\put(32.5,0){\framebox(5,5){}}   \put(35,-1.25){\pos{ct}{$f_3$}}
\put(37.5,2.5){\line(1,0){12.5}}  \put(45,3.5){\pos{cb}{$X_4$}}
\end{picture}
\vspace{-4mm}
\caption{\label{fig:BoxesClosed}%
Closing the inner box in \Fig{fig:Boxes}.}
\end{figure}

Closing a box means replacing it by a single factor
that represents the exterior function of the box. 
Thus, closing the inner box in \Fig{fig:Boxes} yields the NFG
in \Fig{fig:BoxesClosed}.
Opening a box means the reverse process of expanding a factor
into an NFG of its own (with the same exterior function).

\begin{trivlist}
\item{\bf Remark. } closing a box (by summing over the internal variables) and 
opening a box do not change the partition function.
\end{trivlist}

For ease of exposition, we consider NFGs with pairwise interactions between 
the variables (but with arbitrary topology).
We demonstrate the dualization procedure by its 
application to the NFG shown in~\Fig{fig:ExFG4Dual},
which shows an edge of the NFG with factor $f_{k,\ell}(\cdot)$ connected to two equality indicator factors.
To obtain the dual NFG, the dualization procedure needs to be applied 
throughout the primal NFG.

The procedure consists of three steps. 
In the first step, we insert an equality indicator factor into every edge, 
as shown in \Fig{fig:ExFG4Dual1}. More precisely, 
we split each edge, say $X_\ell$, into two edges 
$X_\ell$ and $X_\ell'$, 
which we reconnect via an equality indicator factor.
Clearly, the partition function 
remains unchanged (since configurations in which $X_\ell \neq X_\ell'$ 
do not contribute to the partition function).

In the second step, we expand each of the newly inserted 
equality indicator factors into the product of a (scaled) Fourier kernel $\calF$
and a (scaled) inverse Fourier kernel $\calF^\ast$, as
depicted in \Fig{fig:ExFG4Dual2}. 
We assume that all variables take on 
values in a finite set $\calX$. 
Indeed
\begin{IEEEeqnarray}{c}
\calF(x_\ell,\tilde{x}_\ell) = e^{-\mathrm{i}2\pi x_\ell \tilde{x}_\ell /|\calX|},
\end{IEEEeqnarray}
and
\begin{IEEEeqnarray}{c}
\calF^\ast(x_\ell',\tilde{x}_\ell) = e^{\mathrm{i}2\pi x_\ell' \tilde{x}_\ell /|\calX|},
\end{IEEEeqnarray}
where $\mathrm{i}$ is the unit imaginary number~\cite{Brace:1999}.

\begin{figure*}[t!!!]
\begin{minipage}[b]{0.48\textwidth}
\centering
\setlength{\unitlength}{0.9mm}
\begin{picture}(18,32)
%

%
\put(6,12){\framebox(6,6){}}   
\put(12,15){\line(1,0){22}} 
\put(6,15){\line(-1,0){22}} 

\put(-22,12){\framebox(6,6){=}}         
\put(34,12){\framebox(6,6){=}}
\put(40,15){\line(1,0){9}} 
\put(40,16){\line(1,1){8}} 
\put(40,14){\line(1,-1){8}} 
%
\put(46,19){\pos{cr}{$\vdots$}}
\put(12.25,21){\pos{cr}{$f_{k,\ell}$}}
\put(-1.75,18){\pos{cr}{$X_k$}}
\put(25.25,18.0){\pos{cr}{$X_\ell$}}
\put(-22,15){\line(-1,0){9}} 
\put(-22,16){\line(-1,1){8}} 
\put(-22,14){\line(-1,-1){8}} 
%
\put(-27,19){\pos{cr}{$\vdots$}}
 \put(-17,21){\pos{cr}{$\Phi_{=}$}}
 \put(40,21){\pos{cr}{$\Phi_{=}$}}
\end{picture}
\caption{\label{fig:ExFG4Dual}%
An edge of an NFG with pairwise interactions used to demonstrate the dualization procedure. 
The factor $f_{k,\ell}(\cdot)$ is connected to two equality indicator factors.}
\end{minipage}
\hfill
\begin{minipage}[b]{0.48\textwidth}
\setlength{\unitlength}{0.9mm}
\centering
\begin{picture}(18,32)
\centering
%

%
\put(6,12){\framebox(6,6){}}   
\put(12,15){\line(1,0){9}} 
\put(27,15){\line(1,0){9}} 
\put(6,15){\line(-1,0){9}} 
\put(-9,15){\line(-1,0){9}} 
\put(-24,12){\framebox(6,6){=}}  
\put(-9,12){\framebox(6,6){=}}  
\put(21,12){\framebox(6,6){=}}                
\put(36,12){\framebox(6,6){=}}
\put(42,15){\line(1,0){9}} 
\put(42,16){\line(1,1){8}} 
\put(42,14){\line(1,-1){8}} 
%
\put(48,19){\pos{cr}{$\vdots$}}
\put(12.25,21){\pos{cr}{$f_{k,\ell}$}}
\put(-11,18){\pos{cr}{$X'_k$}}
\put(4,18){\pos{cr}{$X_k$}}
\put(19,18.0){\pos{cr}{$X_\ell$}}
\put(34,18.0){\pos{cr}{$X'_\ell$}}
\put(-24,15){\line(-1,0){9}} 
\put(-24,16){\line(-1,1){8}} 
\put(-24,14){\line(-1,-1){8}} 
\put(-29,19){\pos{cr}{$\vdots$}}
%

%
%

      
\end{picture}
\caption{\label{fig:ExFG4Dual1}%
Inserting an equality indicator factor on every edge of the NFG in Fig.~\ref{fig:ExFG4Dual}. The partition function 
remains unchanged.}
\end{minipage}
\end{figure*}
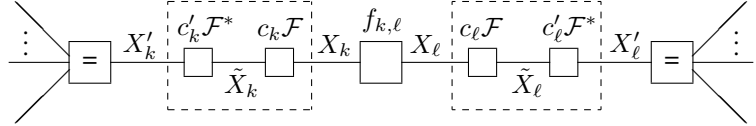
\begin{figure*}
\setlength{\unitlength}{0.9mm}
\centering
\begin{picture}(18,30)
\put(12.25,21){\pos{cr}{$f_{k,\ell}$}}
\put(6,12){\framebox(6,6){}}   
\put(12,15){\line(1,0){10}} 
\put(22,13){\framebox(4,4){}}   
\put(26,15){\line(1,0){8}}       
\put(34,13){\framebox(4,4){}}  
\put(38,15){\line(1,0){11}}     
\put(49,12){\framebox(6,6){=}}  
\put(55,15){\line(1,0){9}} 
\put(55,16){\line(1,1){8}} 
\put(55,14){\line(1,-1){8}} 
%
\put(62,19){\pos{cr}{$\vdots$}}
\put(6,15){\line(-1,0){10}} 
\put(-8,13){\framebox(4,4){}}
\put(-8,15){\line(-1,0){8}} 
\put(-20,13){\framebox(4,4){}} 
\put(-20,15){\line(-1,0){11}}
\put(-37,12){\framebox(6,6){=}}       
\put(27,20){\pos{cr}{$c_{\ell}\calF$}}
\put(41,20){\pos{cr}{$c'_{\ell}\calF^\ast$}}
\put(19.5,8){\dashbox(22,16){}}
\put(18,17.5){\pos{cr}{$X_{\ell}$}}
\put(47.5,17.5){\pos{cr}{$X'_{\ell}$}}
\put(33,12){\pos{cr}{$\tilde X_{\ell}$}}
\put(-2.5,20){\pos{cr}{$c_k\calF$}}
\put(-12.5,20){\pos{cr}{$c'_k\calF^\ast$}}
\put(-22.5,8){\dashbox(21.25,16){}}
\put(-37,15){\line(-1,0){9}} 
\put(-37,16){\line(-1,1){8}} 
\put(-37,14){\line(-1,-1){8}} 
\put(-43,19){\pos{cr}{$\vdots$}}
\put(4.75,17.5){\pos{cr}{$X_k$}}
\put(-24,17.5){\pos{cr}{$X'_k$}}
\put(-9,12){\pos{cr}{$\tilde X_k$}}
\end{picture}
%
\caption{\label{fig:ExFG4Dual2}%
Factoring each equality indicator factor in Fig.~\ref{fig:ExFG4Dual1} into a Fourier transform and an inverse Fourier 
transform.}
\end{figure*}
\vspace{7mm}
\begin{figure*}
\setlength{\unitlength}{0.88mm}
\centering
\begin{picture}(18,30)
\put(12.25,21){\pos{cr}{$f_{k,\ell}$}}
\put(6,12){\framebox(6,6){}}   
\put(12,15){\line(1,0){10}} 
\put(22,13){\framebox(4,4){}}   
\put(26,15){\line(1,0){16}}       
\put(42,13){\framebox(4,4){}}  
\put(46,15){\line(1,0){4}}     
\put(50,12){\framebox(6,6){=}}  
\put(56,15){\line(1,0){6}} 
\put(62,13){\framebox(4,4){}} 
\put(66,15){\line(1,0){8}} 
\put(56,16){\line(1,1){6}} 
\put(62,20.75){\framebox(4,4){}} 
\put(65,24.8){\line(1,1){6}}
\put(56,14){\line(1,-1){6}} 
\put(62,5.25){\framebox(4,4){}} 
\put(65,5){\line(1,-1){6}}
\put(69,21){\pos{cr}{$\vdots$}}
\put(6,15){\line(-1,0){10}} 
\put(-8,13){\framebox(4,4){}}
\put(-8,15){\line(-1,0){16}} 
\put(-28,13){\framebox(4,4){}} 
\put(-28,15){\line(-1,0){4}}
\put(-38,12){\framebox(6,6){=}}       
\put(27,20){\pos{cr}{$c_{\ell}\calF$}}
\put(49,20){\pos{cr}{$c'_{\ell}\calF^\ast$}}
%
%
\put(36,12){\pos{cr}{$\tilde X_{\ell}$}}
\put(-2.3,20){\pos{cr}{$c_k\calF$}}
\put(-21,20){\pos{cr}{$c'_k\calF^\ast$}}
%
%
\put(-38,15){\line(-1,0){6}}
\put(-48,13){\framebox(4,4){}} 
\put(-48,15){\line(-1,0){8}}
\put(-38,16){\line(-1,1){6}}
\put(-48,20.75){\framebox(4,4){}} 
\put(-47,25){\line(-1,1){6}} 
\put(-38,14){\line(-1,-1){6}} 
\put(-48,5.5){\framebox(4,4){}} 
\put(-47.5,5.5){\line(-1,-1){6}} 
\put(-50,21){\pos{cr}{$\vdots$}}
%
\put(-12,12){\pos{cr}{$\tilde X_k$}}
\put(-10,7){\dashbox(38,18){}}
\put(39,3){\dashbox(32,24){}}
\put(-52,3){\dashbox(32,24){}}
\end{picture}
\vspace{2mm}
\caption{\label{fig:ExFG4Dual3}%
Regrouping the factors in~Fig.~\ref{fig:ExFG4Dual2} and closing the dashed boxes yields the dual NFG.}
\end{figure*}
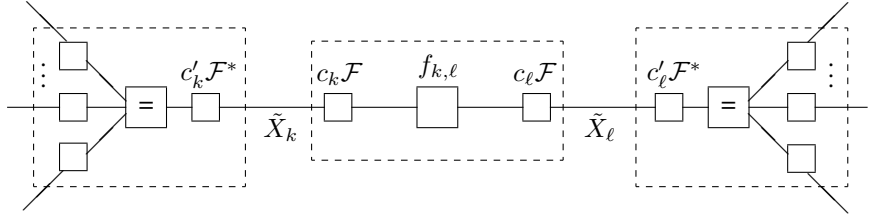
\vspace{1mm}
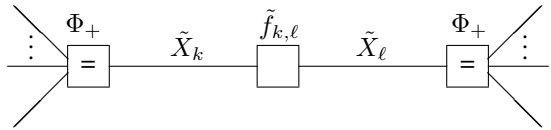
\begin{figure*}
\setlength{\unitlength}{0.9mm}
\centering
\begin{picture}(18,30)
\put(6,12){\framebox(6,6){}}   
\put(12,15){\line(1,0){22}} 
\put(6,15){\line(-1,0){22}} 

\put(-22,12){\framebox(6,6){=}}         
\put(34,12){\framebox(6,6){=}}
\put(40,15){\line(1,0){9}} 
\put(40,16){\line(1,1){8}} 
\put(40,14){\line(1,-1){8}} 
%
\put(46,19){\pos{cr}{$\vdots$}}
\put(12.25,21){\pos{cr}{$\tilde f_{k,\ell}$}}
\put(-1.75,18){\pos{cr}{$\tilde X_k$}}
\put(25.25,18.0){\pos{cr}{$\tilde X_\ell$}}
\put(-22,15){\line(-1,0){9}} 
\put(-22,16){\line(-1,1){8}} 
\put(-22,14){\line(-1,-1){8}} 
%
\put(-27,19){\pos{cr}{$\vdots$}}
 \put(-17,21){\pos{cr}{$\Phi_{+}$}}
 \put(40,21){\pos{cr}{$\Phi_{+}$}}
\end{picture}
\caption{\label{fig:ExFG4Dual4}%
The corresponding edge in the dual of the NFG in Fig.~\ref{fig:ExFG4Dual}.}
\end{figure*}

The 
exterior function of the right dashed box in \Fig{fig:ExFG4Dual2} is
\begin{equation}
\sum_{\tilde{x}_\ell} c_\ell\calF(x_\ell,\tilde{x}_\ell) c'_\ell \calF^\ast(x_\ell',\tilde{x}_\ell)
= c_\ell c'_\ell |\calX|\cdot\delta( x_\ell - x'_\ell)
\end{equation}
and so forth.

An obvious choice for the constants $c$ and $c'$ is such that 
\begin{equation} \label{eqn:AppDualObviousConstants}
c_\ell c'_\ell |\calX| = 1.
\end{equation}

%

In the third step, we regroup the factors as illustrated in \Fig{fig:ExFG4Dual3}.
Closing the dashed boxes in \Fig{fig:ExFG4Dual3} yields 
the dual NFG in \Fig{fig:ExFG4Dual4}, where the factor $f_{k,\ell}(\cdot)$ is
replaced by its Fourier transform $\tilde f_{k,\ell}(\cdot)$, and the equality indicator factors 
$\Phi_{=}(\cdot)$ are replaced by their inverse Fourier transforms, which are 
zero-sum indicator factors $\Phi_{+}(\cdot)$  -- up to scale. 

As in the rest of this paper, we choose the scale factors $c_\ell$ and $c'_\ell$ as
\begin{equation} \label{eqn:AppDualObviousConstants2}
c_\ell = c'_\ell = \frac{1}{|\calX|^{1/2}}\cdot
\end{equation} 

With this choice, the Fourier transform of~(\ref{eqn:IsingA}) is indeed
equal to~(\ref{eqn:IsingKernelD}), and the inverse Fourier transform of an equality indicator 
factor with degree $d$
is
\begin{equation}
\frac{1}{|\calX|^{d/2}}\sum_{x'_{\ell_1}, \ldots, x'_{\ell_d}}\Phi_{=}(x'_{\ell_1}, \ldots, x'_{\ell_d}) \prod_{i = 1}^d \calF^\ast(x'_{\ell_i},\tilde{x}_{\ell_i}),
\end{equation}
which is easily verified to be
\begin{equation}
|\calX|^{1-\tfrac{d}{2}}\cdot\Phi_{+}(\tilde x_{\ell_1}, \ldots, \tilde x_{\ell_d}).
\end{equation}

The (global) scale factor $\alpha(G)$ can then be computed by multiplying all the 
local scale factors as
\begin{IEEEeqnarray}{r;C;l}
\alpha(G) & = & \prod_{i = 1}^N |\calX|^{\tfrac{d_i}{2} - 1}, \\
               & = & |\calX|^{\tfrac{1}{2}\sum_{i=1}^N d_i - |V|}, \\
               & = & |\calX|^{|E| - |V|},
\end{IEEEeqnarray}
where $d_i$ denotes the degree of the $i$-th equality indicator factor, $|V|$ is the number of vertices (which is equal to $N$), 
and $|E|$ denotes the number of edges. 

It should be emphasized that i) the scale factors $c_\ell$ and $c'_\ell$ 
(which were introduced in the second step) 
can be chosen differently,
with a corresponding effect on $\alpha(G)$, and ii) the sequence of $\calF$ and $\calF^\ast$ on every edge 
is arbitrary; but the choice will affect the resulting dual NFG.

%

\section*{Acknowledgements}

The author would like to thank Hans-Andrea Loeliger for his help and advice in providng the Appnedix.
The author also wishes to thank David Forney, Pascal Vontobel, Christoph Pfister, Justin Dauwels, Stefan Moser, and 
 Vicen\c{c} G\'{o}mez for their comments on an earlier version of this paper. 


%
%
%

\newcommand{\IT}{IEEE Trans.\ on Information Theory}
\newcommand{\CASI}{IEEE Trans.\ Circuits \& Systems~I}
\newcommand{\COM}{IEEE Trans.\ Comm.}
\newcommand{\COMLet}{IEEE Commun.\ Lett.}
\newcommand{\COMMag}{IEEE Communications Mag.}
\newcommand{\ETT}{Europ.\ Trans.\ Telecomm.}
\newcommand{\SPMag}{IEEE Signal Proc.\ Mag.}
\newcommand{\ProcIEEE}{Proceedings of the IEEE}

\end{document}